\documentclass[pra,twocolumn,nofootinbib,floatfix]{revtex4-1}

\usepackage{amssymb,amsfonts,amsmath,amsthm}
\usepackage{graphicx}

\usepackage{enumerate}
\usepackage{epstopdf}
\usepackage{color}
\usepackage{euscript}
\usepackage{float}
\usepackage{latexsym}
\usepackage{bm} 
\usepackage{mathrsfs}
\usepackage[colorlinks=true]{hyperref}

\newcommand{\bi}[1]{\textbf{\textit{#1}}}
\newcommand{\bra}[1]{\langle {#1} \vert}
\newcommand{\bbra}[1]{\langle\!\langle {#1} \vert}
\newcommand{\ket}[1]{\vert {#1} \rangle}
\newcommand{\kket}[1]{\vert {#1} \rangle\!\rangle}
\newcommand{\braket}[2]{\langle {#1} \vert {#2} \rangle}
\newcommand{\proj}[1]{\vert {#1} \rangle\!\langle {#1} \vert}
\newcommand{\ketbra}[2]{\vert {#1} \rangle\!\langle {#2} \vert}

\newcommand{\kketbra}[2]{\vert {#1} \rangle\!\rangle\!\langle\!\langle {#2} \vert}
\newcommand{\pproj}[1]{\vert {#1} \rangle\!\rangle\!\langle\!\langle {#1} \vert}
\newcommand{\abs}[1]{\left| {#1} \right|}
\newcommand{\norm}[1]{\abs{\abs{#1}}}
\newcommand{\Tr}[0]{{\mathrm{Tr}}}
\newcommand{\tr}[1]{\text{Tr}\left[ {#1} \right]}
\newcommand{\partialtr}[2]{\text{Tr}_{#1}\left( {#2} \right)}

\newcommand{\blank}[0]{\hspace{3.5mm}}
\newcommand{\acal}[0]{{\mathcal A}}
\newcommand{\bcal}[0]{{\mathcal B}}
\newcommand{\ccal}[0]{{\mathcal C}}
\newcommand{\dcal}[0]{{\mathcal D}}
\newcommand{\hcal}[0]{{\mathcal H}}
\newcommand{\kcal}[0]{{\mathcal K}}
\newcommand{\jcal}[0]{{\mathcal J}}
\newcommand{\ucal}[0]{{\mathcal U}}
\newcommand{\ncal}[0]{{\mathcal N}}
\newcommand{\rcal}[0]{{\mathcal R}}

\newcommand{\ave}[2]{{\langle {#1} \rangle_{#2}}} 

\newcommand{\cont}[1]{{\mathcal C}_{#1}}

\newtheorem{axm}{Axiom}

\begin{document}

\title{Controlled quantum operations and combs, and their applications to \\
universal controllization of divisible unitary operations}

\author{Qingxiuxiong Dong}
 \email{qingxiuxiong.dong@gmail.com}
 \affiliation{Graduate School of Science, The University of Tokyo, Tokyo 113-0033 Japan}
\author{Shojun Nakayama}
 \affiliation{Graduate School of Science, The University of Tokyo, Tokyo 113-0033 Japan}
\author{Akihito Soeda}
 \email{soeda@phys.s.u-tokyo.ac.jp}
 \affiliation{Graduate School of Science, The University of Tokyo, Tokyo 113-0033 Japan}
\author{Mio Murao}
 \email{murao@phys.s.u-tokyo.ac.jp}
 \affiliation{Graduate School of Science, The University of Tokyo, Tokyo 113-0033 Japan}

\begin{abstract}
Unitary operations are a fundamental component of quantum algorithms, but they seem to be far more useful if given with a ``quantum control'' as a controlled unitary operation.
However, quantum operations are not limited to unitary operations.
Nevertheless, it is not a priori clear if a controlled form of these general deterministic quantum operations can be well-defined.
To provide a novel tool in the toolbox for quantum programming, we propose a mathematically consistent definition of a controlled form of deterministic but non-unitary quantum operations and, more generally, of quantum combs.
We propose a ``neutralization'' comb, which transforms a set of input quantum operations to the identity operation,
and study its controlled form based on our definition.
We propose two new quantum algorithms for universal controllization of divisible unitary operations utilizing the most coherently controlled neutralization combs.
\end{abstract}

\date{\today}
\maketitle

\section{Introduction}

Conditional operations describe a branch of operations depending on the input and are fundamental elements in classical and quantum computation~\cite{nielsenchuang,wilde,watrous}.
In classical computation, the simplest conditional operation can be the one taking one bit as a control bit and applying an operation on a target system if the control bit is 1, otherwise leaving the target system unchanged. 
A quantum version of the conditional operation is the \emph{controlled unitary operation},
which takes one qubit as a control qubit, and  it applies a unitary  operation on a target system if the  input state of the control qubit is $\ket{1}$ and  applies the identity  operation if the input state of the control qubit is $\ket{0}$. 
In quantum mechanics, an input state of the control qubit can be taken as an arbitrary superposition of $\ket{0}$ and $\ket{1}$.   
For such an input control state in superposition of $\ket{0}$ and $\ket{1}$, the output state of the quantum version of the conditional operation should be also a coherent superposition of the two branched output states. 
Controlled unitary operations are widely used in quantum computation, for example, Kitaev's phase estimation algorithm~\cite{phase_estimation} and  deterministic quantum computation with one clean qubit  (DQC1)~\cite{dqc1}.

Recently, the effects of a quantum switch~\cite{indefinite1},  %
transforming quantum operations into a channel which coherently executes the input quantum operations in all possible causal orders,
for general quantum operations have been analyzed. 
It is reported that the generalized quantum switch enhances the communication capacity of the input channels, including the completely depolarizing channel~\cite{indefinite2,indefinite3,indefinite_e1}. 
While some allude the enhancement to the indefinitely causally ordered aspect of the quantum switch, others claim that such a phenomenon can happen in systems exploiting coherently controlled quantum operations without causally indefinite elements~\cite{controlchannel1,controlchannel2}.   
Properties of the controlled quantum operations for general deterministic quantum operations depend on the definition, but what should be an ``appropriate'' definition for controlled quantum operations is yet well established.

Quantum algorithms are not necessarily consisting of unitary operations.
There are quantum algorithms utilizing quantum measurements to induce state transformation,
such as in quantum metropolis algorithm~\cite{qmetropolis} and measurement-based quantum computation~\cite{mbqc}.
Quantum combs, a higher-order quantum operation universally transforming quantum operations to other quantum operations~\cite{comb1,comb2}, are also proposed, 
and an algorithm for optimal quantum learning and inversion of an unknown unitary operation is described in terms of the quantum comb~\cite{unitarylearning,unitaryinversion}.
Introducing well-defined controlled versions of general quantum operations and higher-order quantum operations is expected to provide a novel tool in the toolbox for quantum programming, in addition to the standard controlled unitary operations~\cite{qprogramming}.

On the other hand, even though a controlled version of operations is well-defined for unitary operations, it does not mean that universal exact controllization, namely, a higher-order quantum operation universally and exactly transforming an arbitrary unitary operation to its controlled version is implementable in quantum computation.
It has been shown that universal exact controllization of unitary operations maintaining full coherence is impossible with a single use of the unitary operation as an oracle~\cite{controllization1,controllization2,controllization3,controllization4},
and recently, this no-go theorem has been relaxed to finite uses and probabilistic case~\cite{controllization_topological}. 
This no-go theorem of universal controllization of a unitary operation restricts quantum programming, since whenever the controlled unitary operation is called, controllization has to be performed based on the description of each unitary operation, not as a universal quantum operation.

In this paper, we seek an ``appropriate'' definition of a controlled general deterministic quantum operation for utilizing such controlled quantum operations in quantum computation by extending the definition of controlled unitary operations.   
We present a definition of controlled quantum operations with different degrees of coherence,
and the coherence is determined by an operator within the linear span of the Kraus operators.
If the Hilbert-Schmidt norm of this operator is 0, it is classically controlled (no coherence) and if it is 1, it is fully coherently controlled as in the case of standard controlled unitary operations,
and controlled quantum operations with intermediate coherence are also included.
For each quantum operation, the maximal Hilbert-Schmidt norm of the operator is determined, and we call the most coherently controlled quantum operation when such an operator is chosen. 

We further extend our definition to a controlled version of quantum combs~\cite{comb1,comb2}.  %
Then we show applications of the controlled quantum comb to achieve universal controllization, universally implementing maximally coherently controlled quantum operations, for {\it divisible} unitary operations by introducing the idea of a neutralization comb that transforms any quantum operation into the identity operation. 
A controlled neutralization comb can perform a transformation from a quantum operation to its controlled version, although the maximal coherence may not be guaranteed in general.

There are several preceding works on controllization of unitary operations by relaxing the situation of the no-go theorem~\cite{controllization1,controllization2,controllization3,controllization4}. 
It has been shown that if the unitary operation is given by a Hamiltonian dynamics, approximate universal controllization is achieved with an arbitrarily small error by increasing the number of the division of the Hamiltonian dynamics using an auxiliary system~\cite{pme}. 
Also, a necessary and sufficient condition for universal controllization for a set of unitary operations is derived in Ref.~\cite{controllization_iff}.
For particular physical systems, implementations of a controlled unitary operation utilizing a system-specific additional degree of freedom are presented.
For example, for optical interferometer systems, implementations of a controlled unitary operation are proposed using the vacuum degree of freedom~\cite{controllization2,controllization3,controllization_vac_e1}.

As applications of our definitions of the controlled general operations and combs to quantum programming, 
we develop two new quantum algorithms for universal coherent controllization of divisible unitary operations utilizing the most coherently controlled neutralization combs.  
Compared to the previously known algorithms, these algorithms are universal, not depending on the implementation systems, and superior in the following points:  
the first one can achieve universal controllization of a unitary operator in an exact manner with only a finite number of division of the unitary operation;  the second one is implemented without any auxiliary system.

This paper is organized as follows.  
In Section~2, we consider the definition of a controlled version of general quantum operations.
In Section~3, we extend the definition of the controlled quantum operations to controlled quantum combs. 
In Section~4, we investigate the relationship between controlled quantum operations and controlled quantum combs, by introducing the neutralization combs. 
We also present two new quantum algorithms for universal controllization of unitary operations by using the fractional power of the unitary operation.

\section{Definition of the controlled quantum operations}\label{sec:controlled_quantum_channels}

\subsection{Review: Representations of quantum operations}

We first summarize two representations of a general deterministic quantum operation~\cite{nielsenchuang,wilde,watrous}.
Consider a quantum system of which Hilbert space is denoted by $\hcal$.
A quantum state of the quantum system is represented as a positive operator $\rho$ on $\hcal$ with unit trace, 
which is referred to as a density operator. 
The action of a quantum operation ${\mathcal A}$ deterministically transforming a quantum state $\rho$ on $\hcal$ into another quantum state $\rho^\prime = \acal[\rho]$ on another system $\kcal$ can be represented as
\begin{align}
\acal[\rho] = \sum_i K_i \rho K_i^\dagger,
\label{KrausRep}
\end{align}
where each element of the set of the operators $\{ K_i \}$ transforms a state in $\hcal$ to $\kcal$ are called {\it Kraus operators} of the quantum operation.   This representation of a quantum operation is referred to as Kraus representation~\cite{kraus}.
The requirement that quantum operations preserve the trace of density operators leads to the condition $\sum_i K_i^\dagger K_i = I$, where $I$ denotes the identity operator on $\hcal$.

The quantum operation for a unitary operation $\ucal$ is represented as  $\ucal[\rho] = U \rho U^\dagger$ using the corresponding unitary operator $U$.
The Kraus representation of a unitary operation consists of a single Kraus operator $U$.
Note that the global phases of Kraus operators do not affect the action of a quantum operation, 
that is, $\{ K_i \}$ and $\{ e^{i\theta_i} K_i \}$ lead to the same quantum operation.  
In general, Kraus operators of a quantum operation are not uniquely determined,
and there exist different sets of Kraus operators that represent the same quantum operation. 

Another commonly used representation for general quantum operations is the Choi representation~\cite{choi,jamiolkowski}.
In the Choi representation, a quantum operation $\acal:  L(\hcal) \rightarrow L(\kcal)$ is represented as a linear operator on $\hcal \otimes \kcal$ called a Choi operator $J_\acal$ defined by 
\begin{align}
J_\acal = ({\rm id} \otimes \acal ) \pproj{I}_{\hcal \hcal},
\label{eq:def_choi_channel}
\end{align}
where $\kket{I} := \sum_{m} \ket{m}\ket{m}$ is an unnormalized vector in a bipartite system $\hcal \otimes \hcal$ with a fixed orthonormal basis $\{ \ket{m} \}$ of $\hcal$.  
When a quantum operation $\acal$ is  given by Eq.~(\ref{KrausRep}), the corresponding Choi operator can be written as
\begin{align}
 J_\acal =\sum_i \pproj{K_i}_{\hcal \kcal}, \label{eq:def_choi}
\end{align} 
where $\kket{K_i} \in \hcal \otimes {\mathcal K}$ is given by
\begin{align}
  \kket{K_i} = \sum_{mn} \bra{m} K_i \ket{n} \cdot \ket{n}\ket{m}.
\end{align}
In contrast to the Kraus operators, the Choi operator does not depend on the choice of Kraus operators and is uniquely determined by $\acal$.

In particular for a unitary operation $\ucal[\rho] = U \rho U^\dagger$, the corresponding Choi operator is given by 
\begin{align}
J_U &=  ({\rm id} \otimes \ucal ) \pproj{I}_{\hcal \hcal} \\
&= (I \otimes U) \pproj{I}_{\hcal \hcal}  (I \otimes U^\dagger) = \pproj{U}_{\hcal \kcal}.
\end{align}
Note that the Choi operator for the identity operation described by the identity operator $I$ on $\hcal$ to $\kcal$ is given by $J_{I} =\pproj{I}$ on $\hcal \otimes \kcal$, while the projector appearing $\pproj{I}$ in Eq.~(\ref{eq:def_choi_channel}) is an operator on  $\hcal \otimes \hcal$.
In the following of this paper, we explicitly specify the Hilbert space of the vectors and operators by the subscripts when it might be confusing, and omit the subscripts if it is trivial from the context for simplicity.

\subsection{Review: Controlled unitary operations}

A controlled unitary operation is the quantum counterpart of a controlled reversible logic gate in classical computation.
Conventionally, a controlled unitary operation is referred to as a quantum operation that coherently applies different unitary operations on a target quantum system depending on the state of an external qubit called a control qubit.
For a $d$-dimensional unitary operation represented by a unitary operator $U : \hcal (=\mathbb{C}^d) \rightarrow \kcal (=\mathbb{C}^d)$, 
the controlled unitary operation $\cont{U}$ can be defined by the corresponding unitary operator
$ C_{U} : \hcal_C \otimes \hcal \rightarrow \kcal_C \otimes \kcal$ with $\hcal_C={\mathbb C}^2$ and $\kcal_C = {\mathbb C}^2$ given as
\begin{align}
C_{U} := \proj{0} \otimes I + \proj{1} \otimes e^{i\theta_U} U, \label{eq:cont_u}
\end{align}
where $\theta_U$ is an arbitrary phase factor depending on the unitary operator $U$, 
and we always assume the existence of a phase factor when we define a controlled unitary operation.
The degree of freedom of the phase factor $\theta_U$ is required in the definition since for unitary operators $U$ and $e^{i\phi} U$ with a global phase $\phi \in \mathbb{R}$ representing the same operation $\ucal$, 
the corresponding controlled unitary operations $\proj{0} \otimes I + \proj{1} \otimes U$ and $\proj{0} \otimes I + \proj{1} \otimes e^{i\phi} U$ are different unitary operations. 
This phase factor should be defined by a choice and cannot be determined just by specifying the unitary operation $\ucal$.
In particular, for two different unitary operator $U$ and $V = e^{i\phi} U$ representing the same unitary operation,
the corresponding phase factor should satisfy $e^{i \theta_U} = e^{i \theta_V} e^{i\phi}$.
Even if we restrict $U$ to be in $\mathrm{SU}(d)$, the degree of freedom of the phase factor $e^{\frac{2 \pi i}{d} }$ remains and we need to specify which global phase to take for defining the controlled unitary operation described by $C_U$.

The state of an additional control qubit system $\hcal_C$ of $C_{U}$ conditions whether the given unitary operation is applied on the target system or not.
Since the Kraus operator of the controlled unitary operation $\ccal_U$ is $C_U$, 
the Choi operator $J_{\ccal_U}$ on $\hcal_C \otimes \kcal_C \otimes \hcal \otimes \kcal$ for $C_U$ is given by
\begin{align}
J_{\ccal_U} & = ( \ket{00}\kket{I} + \ket{11} \kket{ e^{i \theta_U} U} )( \bra{00} \bbra{I} + \bra{11} \bbra{ e^{i \theta_U} U} )  \label{eq:Choicont_u1} \\
& = \ketbra{00}{00} \otimes J_I + \ketbra{11}{11} \otimes J_U \nonumber \\
 &\quad + \ketbra{00}{11} \otimes \kketbra{I}{e^{i\theta_U} U} + \ketbra{11}{00} \otimes \kketbra{e^{i\theta_U} U}{I}
 \label{eq:Choicont_u}
\end{align}
where $\ket{ii} \in \hcal_C \otimes \kcal_C$ for $i, j=0,1$ denotes a vector of the control system.

An important characteristic of the controlled unitary operation  $\ccal_U$ defined by Eq.~(\ref{eq:cont_u}) is that it preserves coherence between the two different conditioned output states. 
It is also possible to define an incoherent version of a controlled unitary operation where the control qubit is first measured and then the unitary operation is applied or not depending on the measurement outcome. 
The Choi operator of such an incoherently controlled unitary operation is given by 
\begin{align}
J_{\ccal_U^\mathrm{cls}} &= \ketbra{00}{00} \otimes J_I + \ketbra{11}{11} \otimes J_U.
\end{align}
We call $J_{\ccal_U^\mathrm{cls}}$ as a classically controlled version of a unitary operation represented by $U$.   It is straightforward to generalize this classically controlled version of a unitary operation into the one for a general quantum operation, namely, 
\begin{align}
J_{\ccal_\acal^\mathrm{cls}} &:= \ketbra{00}{00} \otimes J_I + \ketbra{11}{11} \otimes J_\acal,
 \label{eq:Choiclscont_u}
\end{align}
which is referred to as a classically controlled version of a quantum operation in this paper.

\subsection{Controlled quantum operations based on physical implementations}

We seek an appropriate definition of  controlled quantum operations by generalizing the definition of the controlled unitary operations preserving coherence $J_{\ccal_U}$ instead of the incoherent version $J_{\ccal_U^\mathrm{cls}}$.    In this subsection, we consider two possible generalizations based on two different implementation schemes of controlled unitary operations. It will turn out that both generalizations emerge to the same definition.

\begin{figure}
  \includegraphics[width=0.9\linewidth]{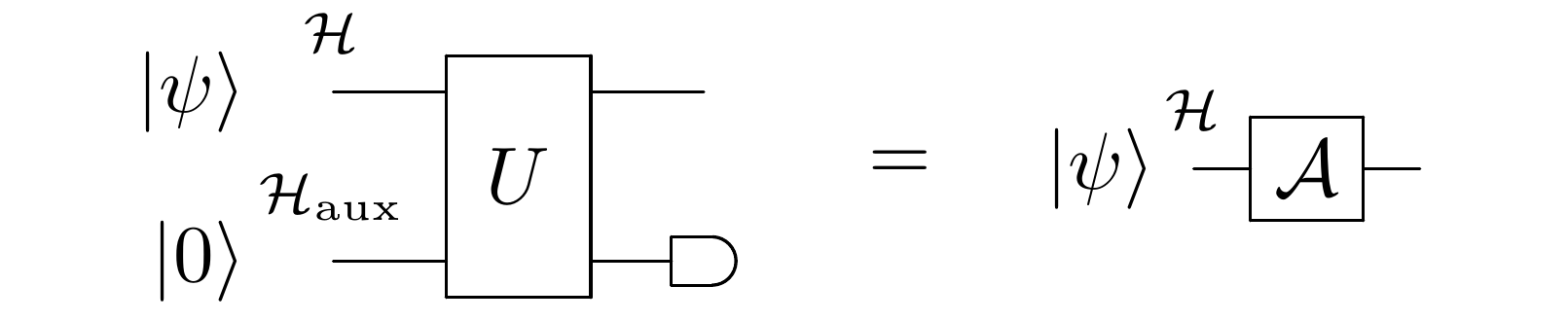}
\caption{
Stinespring representation of a quantum operation.  
A quantum operation $\acal: L(\hcal) \rightarrow L(\kcal)$ can be implemented by adding an auxiliary system $\hcal_{\rm aux}$ in a state $\ket{0}$ to an initial state $\ket{\psi}$ on $\hcal$ and applying a unitary operation $U$ on the joint system $\hcal \otimes \hcal_{\rm aux}$.
The purification of the Kraus representation of $\acal$ can be constructed by using this unitary operator $U$.}
\label{fig:stinespring}
\end{figure}

The first definition of the controlled quantum operation is based on the Stinespring representation~\cite{stinespring} of a quantum operation.  
For a quantum operation ${\mathcal A}$ represented by the Kraus operators given by $\{ K_ i\}_{i=1}^n$, it is always possible to define a unitary operator $U$ on an extended quantum system $\hcal \otimes \hcal_{\rm aux}$ by adding an auxiliary system $\hcal_{\rm aux}={\mathbb C}^{n+1}$ satisfying 
\begin{align}
 U \ket{\psi}\ket{0} = \sum_{i=1}^n K_i \ket{\psi}\ket{i}, \label{eq:def_purification}
\end{align}
where $\{ \ket{i} \}_{i=0}^{n}$ is an orthonormal basis of the auxiliary system.  
Note that we take a particular $U$ such that the summation over $i$ starts from 1 instead of 0 in the r.h.s. of  Eq.~\eqref{eq:def_purification} to treat each Kraus operator $K_i$ for $i=1,\ldots,n$ equally.  
This choice is equivalent to taking the Kraus representation $\{ K_ i\}_{i=0}^n$ with $K_0 = 0$. 
We call this $U$ as a purification of the Kraus representation $\{K_i\}$. 
The quantum operation ${\mathcal A}$ can be represented as the reduced dynamics of this unitary operation as  
\begin{align}
 {\mathcal A} ( \proj{\psi} ) = \Tr_{\rm aux} \left[ U \left( \proj{\psi} \otimes \proj{0} \right) U^\dagger \right],
\end{align}
which corresponds to the quantum circuit shown in Fig.~\ref{fig:stinespring}.

By using Eqs.~(\ref{eq:Choicont_u1}) and (\ref{eq:def_purification}), the Choi operator of the corresponding controlled unitary operation $J_{{\mathcal C}_U} $ on $\hcal_C \otimes \kcal_C \otimes \hcal \otimes \kcal \otimes \hcal_{\rm aux} \otimes \kcal_{\rm aux}$ is given by
\begin{align}
 J_{{\mathcal C}_U} =  \Big(\ket{00}\kket{I}\ket{00}+\ket{11} \sum_{i=1}^n \kket{K_i}\ket{0i}\Big)  \nonumber \\
 \Big( \bra{00}\bbra{I}\bra{00} + \bra{11}\sum_{j=1}^n \bbra{K_j}\bra{0j}\Big). \label{eq:Choi-pufiriedchannel}
\end{align} 
where $\ket{mm}\kket{X}\ket{0n}$ is a tensor product of $\ket{mm} \in \hcal_C \otimes \kcal_C$, $\kket{X} \in \hcal \otimes \kcal $ and $\ket{0n} \in \hcal_{\rm aux} \otimes \kcal_{\rm aux}$. 
We omit the global phase dependence in Eq.~(\ref{eq:Choi-pufiriedchannel}) for simplicity, since it can be absorbed in the notation of $\{ K_i \}$ by choosing the set of the Kraus operators including the choice of the phase factor.  In the rest of this paper, we take this  notation unless it is necessary to explicitly specify the global phase factor.
By tracing out the auxiliary system $\hcal_{\rm aux} \otimes \kcal_{\rm aux}$, the Choi operator of the reduced dynamics is obtained as
\begin{align}
 \Tr_{\rm aux} ( J_{{\mathcal C}_U} ) =  \proj{00} \otimes J_I + \proj{11} \otimes  J_{\mathcal A}.
\end{align} 
Clearly, this is the classically controlled version of a quantum operation ${\mathcal A}$ defined by Eq.~(\ref{eq:Choiclscont_u}).    Even when ${\mathcal A}$ is a unitary operation whose Kraus operator is given by a single element set  $\{K_1 = V\}$ of a unitary operator $V$, the construction of the Choi operator based on the purification given by Eq.~(\ref{eq:def_purification}) derives $J_{\ccal_U^\mathrm{cls}}$ instead of $J_{{\mathcal C}_U}$ preserving coherence.

\begin{figure}
  \includegraphics[width=0.9\linewidth]{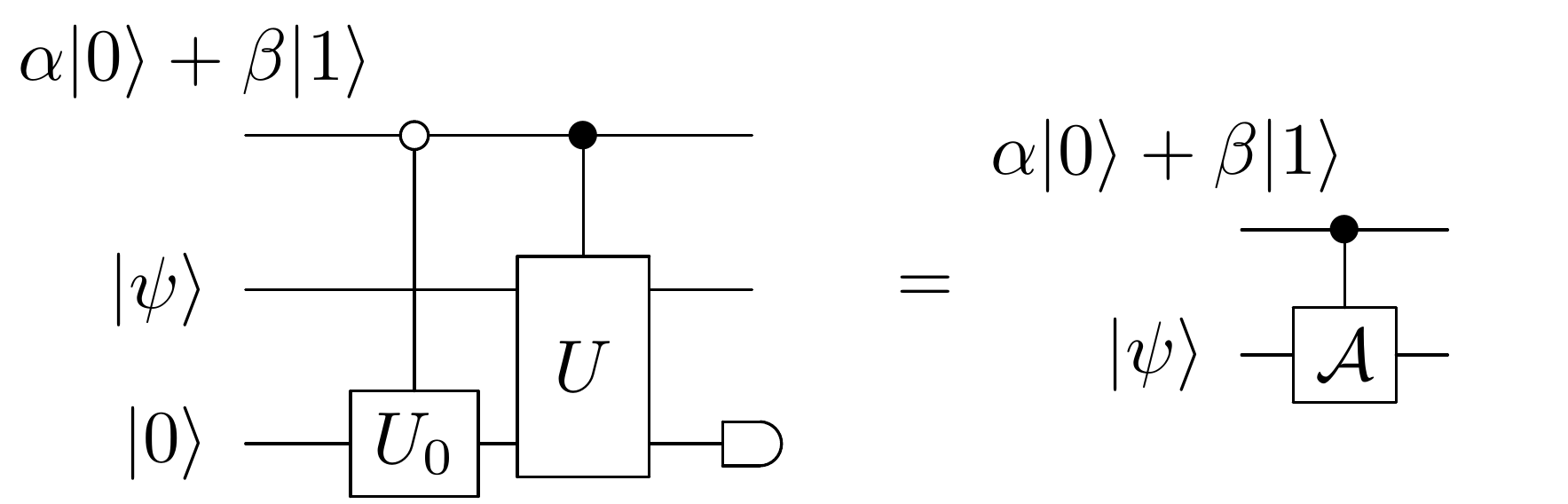}
\caption{A definition of controlled quantum operation based on the Stinespring representation. 
There is an extra degree of freedom by introducing $U_0$ applied to the auxiliary system depending on the state of the control qubit, which can be regarded as a purification of the identity operation.
Given a fixed $U$, controlled quantum operations with all possible degrees of coherence for a given quantum operation $\acal$ are implementable by only changing $U_0$.
}
\label{fig:stinespring2}
\end{figure}

The loss of coherence in this purification originates from ignoring the freedom in the purification of the identity operation applied in the case that the control qubit is $\ket{0}$.
In other words, there is an asymmetry that the identity operation is implemented without purification while ${\mathcal A}$ is. 
The general form of the Kraus representation of the identity operation is given as $\{ K_i = \alpha_i I \}$ satisfying $\sum_i \abs{\alpha_i}^2 = 1$.
The corresponding purification $U_0$ of this Kraus representation of the identity operation is given as
\begin{align}
 U_0 \ket{\psi} \ket{0} = \ket{\psi} \sum_{i=1}^n \alpha_i\ket{i}. 
\end{align}
Note that, $U_0$ is a unitary operator acting nontrivially only on the auxiliary system $\hcal_{\rm aux}$.
We consider that $U_0$ is applied when the control qubit is $\ket{0}$ instead of $I$ on $\hcal \otimes \hcal_{\rm aux}$ in the controlled quantum operation.
Then the corresponding unitary operator of the controlled operation ${\mathcal C}_{U,U_0}$ is 
\begin{align}
 {\mathcal C}_{U,U_0} = \proj{0} \otimes U_0 + \proj{1} \otimes U.
\end{align}
The corresponding Choi operator is given as
\begin{align}
 J_{{\mathcal C}_{U,U_0}} =&  \sum_{i,j=1}^n \Big(\alpha_i \ket{00}\kket{I}+\ket{11}\kket{K_i}\Big)\cdot \notag \\
&\quad \Big( \alpha_j^* \bra{00}\bbra{I} + \bra{11} \bbra{K_j}\Big)\otimes \ketbra{0i}{0j},
\end{align}
By tracing out the auxiliary system $\hcal_{\rm aux} \otimes \kcal_{\rm aux}$, we obtain
\begin{align}
 &\partialtr{\rm aux}{J_{{\mathcal C}_{U,U_0}}} =  \notag \\
 &\sum_{i} \Big(\alpha_i \ket{00}\kket{I}+\ket{11}\kket{K_i}\Big) \Big( \alpha_i^\ast \bra{00}\bbra{I} + \bra{11} \bbra{K_i}\Big). \label{eq:def_cc1}
\end{align}
The corresponding quantum circuit is shown in Fig.~\ref{fig:stinespring2}.
We take the definition given by Eq.~\eqref{eq:def_cc1} as the first definition of a controlled quantum operation.
For a given quantum operation, controlled quantum operations with different degrees of coherence can be defined by changing the set of Kraus operators $\{ K_ i\}$ and coefficients $\{ \alpha_i \}$.

In the appendix of Ref.~\cite{controlchannel1}, a definition of a controlled quantum operation is introduced in terms of purification with an environment.  
They obtain a similar representation to ours where $\alpha_i = \bra{i} U_0 \ket{0}$ in Eq.~(\ref{eq:def_cc1}) is given by $\braket{i}{\varepsilon_0}$ with an initial state of the environment $\ket{\varepsilon_0}$ in their definition.
The main difference between the definition of Ref.~\cite{controlchannel1} and our definition is that we  explicitly choose a certain type of Kraus operators $\{ K_ i\}_{i=0}^n$ satisfying $K_0 = 0$.
By this choice of the Kraus operators and the corresponding purification $U$,
the quantum circuit shown in Fig.~\ref{fig:stinespring2} can implement controlled quantum operations with all possible degrees of coherence by just choosing the coefficients $\{ \alpha_i \}$ or equivalently $U_0$.
We will  analyze the point at the end of Sec.~\ref{sec:axiomatic_controlled_channel}.

\begin{figure}
\centering
\includegraphics[width=\hsize]{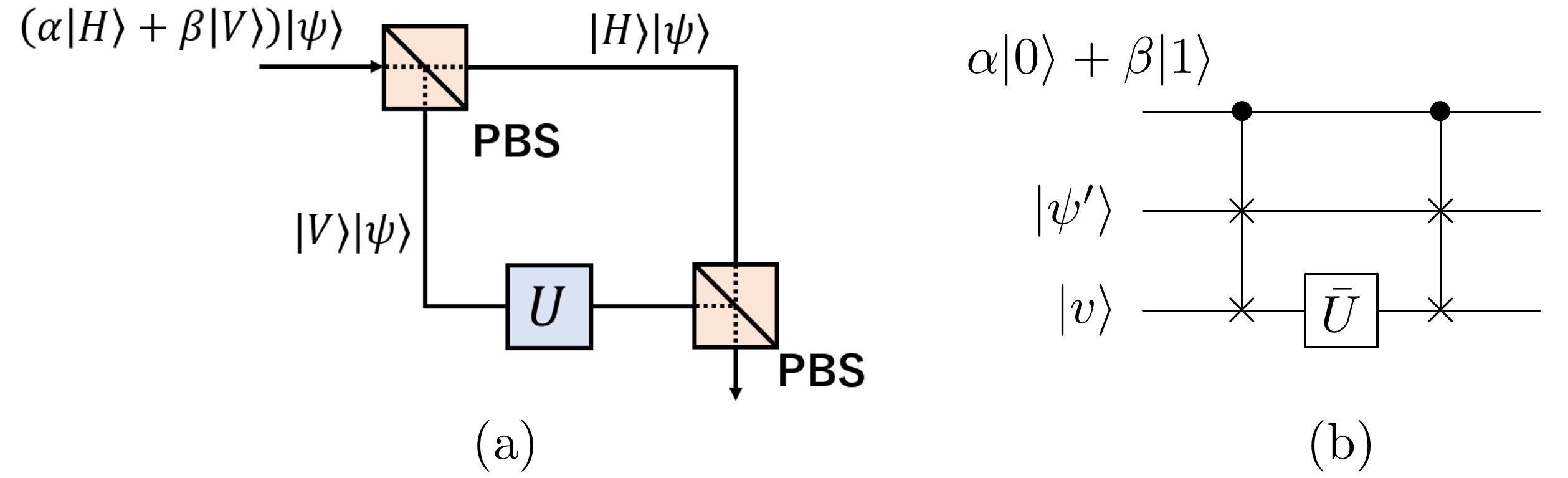} 
\caption{(a) An implementation of a controlled unitary operation in an optical interferometer~\cite{controllization2,controllization3}.
The optical element PBS denotes a polarization beam splitter, and the polarization $\ket{H}$ and $\ket{V}$ control the routing of the optical path.
The lower path has an optical element acting on the additional degrees of freedom, which corresponds to the unitary operation $U$ acting on the target system.
(b) An equivalent quantum circuit to the optical interferometer by introducing the vacuum state $\ket{v}$.
The control qubit $\ket{0}$ and $\ket{1}$ correspond to the polarization $\ket{H}$ and $\ket{V}$.
The target state $\ket{\psi'}$ is a quantum state embedded in a one-dimension larger Hilbert space $\ket{\psi'} \in \hcal \oplus {\mathbb C}$ that is equivalent to the original target state $\ket{\psi} \in \hcal$ as $\ket{\psi'} = \ket{\psi} + 0 \ket{v}$.
The auxiliary state is given by the vacuum state $\ket{v} \in \hcal \oplus {\mathbb C}$,
and the unitary operation $\bar{U} = U \oplus \proj{v} \in L (\hcal \oplus {\mathbb C})$ does not change the vacuum state $\ket{v}$.}\label{fig:optical}
\end{figure}

The second definition of a controlled quantum operation is to use an additional dimension,
based on the implementation of a controlled unitary operation in the optical interferometer system~\cite{controllization2,controllization3,shannon_path,controllization_vac_e1, qs_ind_exp, qs_comm_theory1}.
Consider a composite of quantum states of the control qubit $\alpha\ket{0} + \beta\ket{1}$ and the target state $\ket{\psi}$.
We assume that the control qubit and the target state are encoded into a single photon.
That is, the control qubit is encoded into the polarization of a photon as $\alpha\ket{H} + \beta\ket{V}$, where $\ket{H}$ and $\ket{V}$ denote the horizontal and the vertical polarization,
and the target state is encoded into other degrees of freedom of the same photon such as the orbital angular momentum or the transverse spatial mode, which is represented by the Hilbert space $\hcal$.
A unitary operation represented by $U \in {\mathcal L}(\hcal)$ can be realized by an optical element 
which acts on the additional degrees of freedom but not the polarization, 
and the corresponding controlled unitary operation ${\mathcal C}_U$ can be implemented with the optical interferometer shown in Fig.~\ref{fig:optical}a.
The polarization of the photon controls its path via polarization beam splitters, and the optical elements corresponding to $U$ is placed in the lower path.
If the polarization of the photon is in $\ket{V}$, the photon passes through the lower path and $U$ is applied on the target state $\ket{\psi}$.
If the polarization of the photon is in $\ket{H}$, the photon passes through the upper path, 
and the vacuum state passes through the optical elements corresponding to $U$ which remains to be the vacuum state.
Thus, the resulting quantum state is given by $\alpha\ket{H}\ket{\psi} + \beta\ket{V} U\ket{\psi}$, and the action of the controlled unitary operation is obtained.
By considering the vacuum state $\ket{v}$, which is ignored in the formulation of optical elements, 
a unitary operation $U$ on the Hilbert space $\hcal$ can be regarded as 
a unitary operation $\bar{U}$ embedded into a one-dimension larger Hilbert space $\hcal \oplus {\mathbb C}$ as $\bar{U} = U \oplus \proj{v}$,
where $\oplus$ denotes the direct sum.
The optical interferometer shown in Fig.~\ref{fig:optical}a with the unitary operation $U$ 
can be regarded as the quantum circuit shown in Fig.~\ref{fig:optical}b with the unitary operation $\bar{U}$.

An embedded unitary operation $\bar{U}$ is transformed to the corresponding controlled unitary operation ${\mathcal C}_{\bar{U}}$ by the optical interferometer shown in Fig.~\ref{fig:optical}.  This transformation is represented by the following function $f$, namely,
\begin{align}
 f(J_{\bar{U}}) %
 &=\proj{00} \otimes J_{\bar{I}} + \proj{11} \otimes J_{\bar{U}} \\
 &\quad + \ketbra{00}{11} \otimes \kketbra{\bar{I}}{\bar{U}} + h.c. \\
 &=\proj{00} \otimes J_{\bar{I}} + \proj{11} \otimes J_{\bar{U}} \\
 &\quad + \ketbra{00}{11} \otimes \kket{\bar{I}} (\bra{vv} J_{\bar{U}} ) + h.c.,
\end{align}
where $\kket{\bar{U}} = \kket{U} + \ket{vv}$ by definition of the embedded space, and the second equality holds because $\langle vv \kket{\bar{U}} = 1$.
Note that the function $f$ only depends on $J_{\bar{U}}$, which is uniquely determined for a unitary operation.

This optical interferometer implementation for a controlled unitary operation can also be extended for general quantum operations.
A quantum operation ${\mathcal A}$ can be extended to $\bar{\acal}$ of a larger dimensional system by extending the Kraus operators as $\bar{K_i}=K_i \oplus \alpha_i \proj{v}$, 
where coefficients $\{\alpha_i \}$ satisfying $\sum_i | \alpha_i |^2 = 1$ are necessary so that $\{ \bar{K_i} \}$ is also a Kraus representation of a quantum operation.
For a quantum operation given by the Kraus operators $\{ \bar{K_i} \}$, 
the controlled version given by the optical interferometer shown in Fig.~\ref{fig:optical} is uniquely determined by the Choi operator $f(J_{\bar{\acal}})$ as 
 \begin{align}
 f({J}_{\bar{\mathcal A}}) 
&=\proj{00} \otimes J_{\bar{I}} + \proj{11} \otimes J_{\bar{\acal}} \\
 &\quad + \ketbra{00}{11} \otimes \kket{\bar{I}} (\bra{vv} J_{\bar{\acal}} ) + h.c. \\ 
 &=  \sum_{i} \left( \alpha_i \ket{00}\kket{I}+\ket{11}\kket{K_i} + \alpha_i \ket{\xi} \ket{vv} \right) \\ 
&\quad \times ( \alpha_i^\ast \bra{00}\bbra{I} + \bra{11} \bbra{K_i} + \alpha_i^\ast \bra{\xi} \bra{vv} ), \label{eq:def_cc2_raw}
\end{align}
where $\ket{\xi} = \ket{00} + \ket{11}$ is a vector in $\hcal_C \otimes \kcal_C$.
Assuming that the input state of this extended operation does not contain the vacuum state, that is, 
it is orthogonal to $\ket{\psi_{\rm ctrl}}\ket{v}$, where $\ket{\psi_{\rm ctrl}}$ is an arbitrary state of the control qubit,
the third term in each bracket does not affect the result, and Eq.~\eqref{eq:def_cc2_raw} is equivalent to Eq.~\eqref{eq:def_cc1}. 
 
As the two different definitions emerge to identical ones, we take the definition of the controlled quantum operation derived as
\begin{multline}
 J_{{\mathcal C}_{\mathcal A}^{ K_i, \alpha_i }} =  \\
 \sum_{i} \Big(\alpha_i \ket{00}\kket{I}+\ket{11}\kket{K_i}\Big) \Big( \alpha_i^\ast \bra{00}\bbra{I} + \bra{11} \bbra{K_i}\Big). \label{eq:def_cc2}
\end{multline}
We stress that a characteristic property of this definition of a controlled quantum operation $J_{{\mathcal C}_{\mathcal A}^{ K_i, \alpha_i }}$ is that it cannot be uniquely determined by the Choi operator $J_\acal$, but depends on both the choice of the Kraus operators $\{ K_i \}$ and the coefficients $\{ \alpha_i \}$.  This is in contrast to the classical controlled version of a quantum operation $J_{\ccal_\acal^\mathrm{cls}}$, which is uniquely determined for $\acal$ irrespective of the choice of $\{ K_i \}$ and $\{ \alpha_i \}$.

\subsection{Axiomatic definition of the controlled quantum operations}\label{sec:axiomatic_controlled_channel}

In this subsection, we define a controlled quantum operation in an axiomatic manner.
We show that we can derive the definition of the form given in Eq.~\eqref{eq:def_cc2} from a small number of axioms. 
We consider that a controlled quantum operation should satisfy the following three criteria. 
\begin{axm}[Axioms for controlled quantum operations] \label{axm:cc}
The action of a controlled quantum operation of a deterministic quantum operation ${\mathcal A}$ satisfies the following three axioms. 
\begin{enumerate}
 \item If the state of control qubit is $\ket{0}$ or $\ket{1}$, the state of the control qubit remains unchanged after applying the controlled quantum operation.   
 \item If the state of the control qubit is $\ket{0}$, the identity operation is applied on the target system.
 \item If the state of the control qubit is $\ket{1}$, the quantum operation ${\mathcal A}$ is applied on the target system.  
\end{enumerate}
\end{axm}

The form of the controlled quantum operation given by Eq.~\eqref{eq:def_cc2} can be derived from just these three axioms as follows. 
A general form of the Kraus operators of a quantum operation on a composite system consisting of a control system and a target system is written by 
\begin{align} 
 L_i &= \proj{0} \otimes A_i + \ketbra{0}{1} \otimes B_i + \ketbra{1}{0} \otimes C_i + \proj{1} \otimes D_i \notag\\
 &\quad \in L( \hcal_C \otimes \hcal , \kcal_C \otimes \kcal ).
\end{align} 
Due to the first axiom, terms in $L_i$ that change the control qubit state must be zero, that is, that is, $B_i =C_i =0$. 
To satisfy the second axiom, each $A_i$ must be proportional to identity, that is, 
$A_i = \alpha_i I$ with $\sum_i | \alpha_i |^2 = 1$ is required.
The third axiom implies that $\{ D_i \}$ forms a Kraus representation of ${\mathcal A}$. 
Therefore, the Kraus operator of the controlled quantum operation has to be in a form of
\begin{align} 
 L_i = \alpha_i  \proj{0} \otimes I + \proj{1} \otimes K_i, \label{eq:kraus_channel}
\end{align} 
where $\sum_i | \alpha_i |^2 = 1$ and $\{ K_i \}$ is a Kraus representation of $\acal$.
The quantum operation given by the Kraus representation $\{ L_i \}$ is equivalent to that of Eq.~\eqref{eq:def_cc2}.

A controlled quantum operation is characterized by the parameters $\{ \alpha_i \}$ and $\{ K_i \}$,
but not all different combinations of these correspond to all different controlled quantum operations in general, 
namely, these parameters are redundant.
In the following, we provide a parameterization that uniquely determines a controlled quantum operation.
By expanding Eq.~\eqref{eq:def_cc2}, we obtain  
\begin{align}
 J_{ {\mathcal C}_{\mathcal A}^{ K_i, \alpha_i } } &=  J_{ {\mathcal C}_{\mathcal A}^{\rm cls} } \nonumber \\
& + \ketbra{00}{11} \otimes \kketbra{I}{K} + \ketbra{11}{00} \otimes \kketbra{K}{I}, 
\end{align} 
where $J_{ {\mathcal C}_{\mathcal A}^{\rm cls} }$ is a controlled quantum operation of ${\mathcal A}$ without coherence  defined by Eq.~(\ref{eq:Choiclscont_u}) and $K $ is the operator given by $K = \sum_i \alpha_i^\ast K_i$. 
As a Choi operator uniquely determines a quantum operation, the operator $K$ fully specifies one controlled quantum operation of ${\mathcal A}$ without redundancy.
Note that the operator $K$ corresponds to the transformation matrix introduced in Ref.~\cite{controlchannel1}.
In the following, we use the definition of the controlled quantum operation ${\mathcal C}_{\mathcal A}^K$ for $\acal$ with a choice of $K = \sum_i \alpha_i^\ast K_i$ for $\acal$ as 
\begin{align}
 J_{ {\mathcal C}_{\mathcal A}^{ K } } &:=  J_{ {\mathcal C}_{\mathcal A}^{\rm cls} } \nonumber \\
& + \ketbra{00}{11} \otimes \kketbra{I}{K} + \ketbra{11}{00} \otimes \kketbra{K}{I}.  \label{eq:def_cck}
\end{align}

 Now we show that the quantum circuit shown in Fig.~\ref{fig:stinespring2} can implement controlled quantum operations with all possible degrees of coherence for a given quantum operation by choosing only the coefficients $\{ \alpha_i \}$ or equivalently $U_0$.  More precisely, given a fixed set of Kraus operators $\{ K_i \}_{i=0}^{n}$ with $K_0 = 0$,
for any set of the Kraus operators $\{ K'_j \}_{j=0}^{m}$ with $K'_0=0$ and coefficients $\{ \alpha'_j \}_{j=0}^{m}$ with $\sum_j | \alpha'_j |^2 =1$,
we can choose $\{ \alpha_i \}$ with $\sum_i | \alpha_i |^2 =1$ to make the two resulting controlled quantum operations to be equivalent. This property implies that we can represent all possible controlled quantum operations for a given quantum operation by any choice of the Kraus operators with our definition given by Eq.~\eqref{eq:def_cc1}.

Since $\{ K_i \}_{i=0}^{n}$ and $\{ K'_j \}_{j=0}^{m}$ represent the same quantum operation, 
$K'_j = \sum_i u^*_{ji} K_i$ holds with a unitary matrix $( u_{ij} )$~\cite{nielsenchuang}.
Note that if $n \neq m$, we pad with $K_i,K'_j = 0$ to make the number of the Kraus operators to be the same.
In order to implement the controlled quantum operation with $K=\sum_j (\alpha'_j)^* K'_j $,
by considering $\sum_j (\alpha'_j)^* K'_j = \sum_{i,j=0}^{n} (\alpha'_j)^* u^*_{ji} K_i$,
we can choose $\alpha_i = \sum_j \alpha'_j u_{ji}$ when the Kraus operators are given by $\{ K_i \}_{i=0}^{n}$.
For $n < m$, $\sum_{i=0}^n | \alpha_i |^2$ may be smaller than 1, we pad $\{ K_i \}_{i=0}^{n}$ to $\{ K_i \}_{i=0}^{m}$ with $K_i=0$ for $n < i \leq m$.
However, since we assumed $K_0 = 0$, we can re-define the coefficient $ | \alpha_0 |^2$ by $ | \alpha_0 |^2 + \sum_{i=n+1}^m | \alpha_i |^2 $ to satisfy $\sum_{i=0}^n | \alpha_i |^2 = 1$.
Note that the phase of $\alpha_0$ can be chosen arbitrarily as the corresponding Kraus operator is $K_0 = 0$.
For $n \geq m$, $\sum_{i=0}^n | \alpha_i |^2 = 1$ is satisfied by construction.
Thus, for a given set of the Kraus operators $\{ K_i \}_{i=0}^{n}$ with $K_0 = 0$, we can choose only the coefficients $\{ \alpha_i \}$ to define controlled quantum operations with all possible degrees of coherence,
and it also indicates that the quantum circuit shown in Fig.~\ref{fig:stinespring2} can be used to implement  corresponding all possible controlled quantum operations.

\subsection{Most coherently controlled quantum operation}\label{sec:coh_controlled_channel}

The controlled quantum operation defined by ${\mathcal C}_{\mathcal A}^K$ contains different types of controlled quantum operations including the classically controlled version of quantum operations depending on the choice of $K$.
However, in quantum information processing, keeping coherence or superposition of states is important 
and thus we consider how to characterize the most coherently controlled quantum operation in this subsection.

Since we focus on the coherence between the different states of the control qubit,
we investigate the (block) off-diagonal term of the corresponding Choi operator of a controlled quantum operation.
Especially, we analyze the off-diagonal term of the Choi operator indicating coherence of a controlled quantum operation given by
\begin{align}
 \Delta J_{{\mathcal C}_{\mathcal A}^K} = J_{{\mathcal C}_{\mathcal A}^K} - J_{{\mathcal C}_{\mathcal A}^{\rm cls}}.
\end{align}
The trace norm of $\Delta J_{{\mathcal C}_{\mathcal A}^K}$ corresponds to a distance measure between quantum operations~\cite{distance}.
We regard that the controlled quantum operation that has the largest norm of the off-diagonal term,
 which can be also interpreted to be the most distant one from the classically controlled version, as the quantum mechanically most coherent one.
Here $\Delta  J_{{\mathcal C}_{\mathcal A}^K }$ has only two non-zero eigenvalues $\lambda = \pm \sqrt{d} \sqrt{ \tr{K^\dagger K}}$ and the corresponding two eigenstates are given by $(1/\sqrt{d} ) \ket{00} \otimes \kket{I} \pm (1/\sqrt{\Tr K^\dag K}) \ket{11} \otimes \kket{K}$.
Thus, we obtain the Schatten $p$-norm\footnote{The trace norm corresponds to $p=1$.} of this operator $\norm{ X }_p =  \sqrt[p]{\tr{ \abs{X}^p }}$ as
\begin{align}
 \norm{\Delta J_{{\mathcal C}_{\mathcal A}^K }}_p = 2^{\frac{1}{p}} \sqrt{d} \sqrt{ \tr{K^\dagger K} }.   \label{eq:coh} 
\end{align}

According to Eq.~\eqref{eq:coh}, ${\mathcal C}_{\mathcal A}^K$ with maximum quantum coherence (in the sense of the Schatten $p$-norm) is obtained by maximizing the Hilbert-Schmidt norm of $K$.
In order to calculate the Hilbert-Schmidt norm, we introduce the orthogonal Kraus representation $\{ \tilde{K}_j \}$ as follows.
For any quantum operation ${\mathcal A}$, we can take a set of mutually orthogonal Kraus operators $\{ \tilde{K}_j \}_{j=1}^{m}$ 
 satisfying $\tr{\tilde{K_i}^\dagger \tilde{K_j}} = 0$ for all $i \neq j$.
Explicitly, $\{ \tilde{K_j} \}_{j=1}^{m}$ can be obtained by first calculating the Choi operator of ${\mathcal A}$, 
and then performing the spectral decomposition on the Choi operator.
Note that the number of the Kraus operator in the orthogonal Kraus representation $\{ \tilde{K}_j \}_{j=1}^{m}$  satisfies $m \leq n$ where $n$ is the number of the Kraus operators in an arbitrary Kraus representation $\{ K_i \}_{i=1}^{n}$.
We can rewrite $K = \sum_i \alpha_i^\ast K_i$ as $K = \sum_i\beta_i^\ast \tilde{K_i}$ with 
$$ \beta_i = \frac{\tr{K^\dagger \tilde{K_i}}}{\tr{\tilde{K_i}^\dagger \tilde{K_i}}}$$
 by defining $\tilde{K_i} = 0$ for $m <i \leq n$ when $m < n$.
The coefficients $\{ \beta_i \}$ satisfy $\sum_i \abs{\beta_i}^2\leq 1$ and this can be shown as follows.
Since $\{ K_i \}$ and $\{ \tilde{K_j} \}$ represent the same quantum operation, 
$K_i = \sum_{j=1}^{n} u_{ij} \tilde{K_j}$ holds with a unitary matrix $( u_{ij} )$~\cite{nielsenchuang}. 
Then we obtain
\begin{align}
\sum_{j=1}^m \abs{\beta_i}^2 &= \sum_{i,j=1}^n \abs{\alpha_i}^2  \abs{u_{ij}}^2 \delta_{i \leq m} \nonumber \\
&= \sum_{i=1}^m \abs{\alpha_i}^2 \leq \sum_{i=1}^n \abs{\alpha_i}^2 = 1
\end{align}
where $\delta_{i \leq m}$ denotes a step function, namely,  $\delta_{i \leq m} =1$ for ${i \leq m}$ and otherwise $\delta_{i \leq m} =0$.

By using the orthogonal Kraus representation $\{ \tilde{K_i} \}$, the Hilbert-Schmidt norm of $K$ is represented as
\begin{align}
 \tr{K^\dagger K} = \sum_{i} \abs{\beta_i}^2 \tr{\tilde{K_i}^\dagger \tilde{K_i}}, ~ \sum_i \abs{\beta_i}^2 \leq 1. \label{decK} 
\end{align}
We define a subset of the index of the orthogonal Kraus operators $\{ \tilde{K_i} \}$ of a quantum operation ${\mathcal A}$  with the maximum Hilbert-Schmidt norm as
\begin{align}
 A_{\rm max}:= \left\{ i~ \Big\vert~ \forall j, ~ \tr{\tilde{K_i}^\dagger \tilde{K_i}} \geq \tr{\tilde{K_j}^\dagger \tilde{K_j}} \right\}.
\end{align}
It is clear from Eq.~\eqref{decK} that the operator $K$ for the maximum coherence is obtained by appropriately choosing the coefficients $\{ \alpha_i \}$ for the orthogonal Kraus operators with the maximum Hilbert-Schmidt norm as
\begin{align}
 K = \sum_{i \in A_{\rm max}} \alpha_i^\ast \tilde{K_i},~ \sum_i \abs{\alpha_i}^2 = 1.
\end{align}
We can construct an orthogonal Kraus representation of ${\mathcal A}$ which includes $K$ as one of the Kraus operators as $\{ K = K'_1, K'_2, \ldots, K'_m \}$
by choosing $K'_i = \sum_{j=1}^m v_{ij} \tilde{K}_j$ with a unitary matrix $(v_{ij})$ satisfying $v_{1j} = \alpha_j^\ast$.
In other words, $K$ is one of the possible Kraus operators of ${\mathcal A}$ which has the maximum Hilbert-Schmidt norm.   
In the following, we call a controlled quantum operation of ${\mathcal A}$ described with the maximum Hilbert-Schmidt norm of $K$ as the {\it most coherently controlled quantum operation}.
In particular, when the maximal Hilbert-Schmidt norm of $K$ is 1, we refer such a controlled quantum operation to as the {\it fully} coherently controlled quantum operation. 
The controlled unitary operations given by Eq.~(\ref{eq:Choicont_u}) is the fully coherently controlled quantum operation as expected.

Remark that the definition of a controlled quantum operation can be generalized by replacing the identity operation applied when the control qubit state is $\ket{0}$ by another general quantum operation, such as the depolarizing channel.
Such kinds of controlled quantum operations are considered in Ref.~\cite{controlchannel1,controlchannel2}.
However, it is difficult to evaluate the coherence in general for such cases, and it is not clear what can be regarded as  the most coherently controlled quantum operation.
Nevertheless, the most coherently controlled quantum operation can be easily extended if another quantum operation described by a single Kraus operator, e.g., an isometry $V$, is applied when the control qubit state is $\ket{0}$,  instead of the identity operation.
In this case, by replacing the identity operator by the isometry $V$ in Eq.~\eqref{eq:kraus_channel}, all the calculation directly follows, and the eigenvalue to calculate Eq.~\eqref{eq:coh} becomes $\pm \sqrt{\tr{V^\dag V}} \sqrt{\tr{K^\dag K}} $.
Since the trace-preserving condition is given as $V^\dag V = I$ for deterministic quantum operations, 
we obtain the same value of coherence as Eq.~\eqref{eq:coh}.

\subsection{Controlled quantum operations switching between two quantum operations}

In the previous sections, we only considered controlled quantum operations which apply the identity operation when the control qubit is in state $\ket{0}$.
In this subsection, we consider controlled quantum operations which switch between two quantum operations depending on the control qubit,
namely, a controlled quantum operation which applies quantum operation $\acal$ if the control qubit is in $\ket{0}$,
and applies quantum operation $\bcal$ if the control qubit is in $\ket{1}$.
Such a controlled quantum operation can be realized by a concatenation of the controlled version of $\acal$ and $\bcal$ 
(with the first one having inverted control qubit, namely, applying $\acal$ if the control qubit is in $\ket{0}$ and do nothing if in $\ket{1}$).
\begin{axm}
The action of a controlled quantum operation switching between two deterministic quantum operations ${\mathcal A}$ and ${\mathcal B}$ satisfies the following three axioms. 
\begin{enumerate}\label{axm:cc2}
 \item If the state of control qubit is $\ket{0}$ or $\ket{1}$, the state of the control qubit remains unchanged after applying the controlled quantum operation.   
 \item If the state of the control qubit is $\ket{0}$, the quantum operation ${\mathcal A}$ is applied on the target system.
 \item If the state of the control qubit is $\ket{1}$, the quantum operation ${\mathcal B}$ is applied on the target system.  
\end{enumerate}
\end{axm}
In order to avoid confusion, we denote the controlled quantum operation of ${\acal}$ with inverted control qubit as $\bar{\ccal}_\acal$ in the following sections.
That is, $\bar{\ccal}_\acal := {\mathcal X}_c \circ {\ccal}_\acal \circ {\mathcal X}_c$ where ${\mathcal X}_c$ denotes the Pauli $X$ operation (NOT operation) on the control qubit.

The resulting controlled quantum operation of concatenation of the controlled version of $\acal$ and $\bcal$,
i.e., $\bar{\ccal}_{\acal}^{K} \circ {\ccal}_{\bcal}^L = {\ccal}_{\bcal}^L \circ \bar{\ccal}_{\acal}^{K} $, is given by
\begin{align}
&\proj{00} \otimes J_\acal + \proj{11} \otimes J_\bcal \notag \\
&\quad+ \ketbra{00}{11} \otimes \kketbra{K}{L} + \ketbra{11}{00} \otimes \kketbra{L}{K}, \label{eq:def_cc_2op}
\end{align}
where $K$ and $L$ denotes the Kraus operators defining each controlled quantum operation ${\ccal}_{\acal}^{K}$ and ${\ccal}_{\bcal}^{L}$.
However, the concatenation does not define the most general controlled quantum operation between $\acal$ and $\bcal$.
For example, consider the two-dimensional case where both $\acal$ and $\bcal$ are the depolarizing channel $\dcal$.
In this case, the quantum operation $\mathrm{id} \otimes \dcal$ can be considered as a controlled version,
because it satisfies Axiom~\ref{axm:cc2} as it applies $\dcal$ when control qubit is in $\ket{0}$ or $\ket{1}$, and does not change the state of the control qubit.
However, this quantum operation cannot be realized by a concatenation given by Eq.~\eqref{eq:def_cc_2op}.
That is, the Choi operator of $\mathrm{id} \otimes \dcal$ is given by
\begin{align}
(\proj{00} + \proj{11} + \ketbra{00}{11} + \ketbra{11}{00}) \otimes  \frac{I}{2},
\end{align}
thus the off-diagonal term corresponding to $\ketbra{00}{11}$ is given by $\frac{I}{2}$.
On the other hand,  the off-diagonal term corresponding to $\ketbra{00}{11}$ for the concatenation Eq.~\eqref{eq:def_cc_2op} is given by $\kketbra{K}{L}$.
Since the Hilbert-Schmidt norm of $K$ and $L$ is bounded by $1/2$ for the two-dimensional depolarizing channel,
the inequality $| \Tr \kketbra{K}{L} | = | \Tr L^\dag K | \leq \sqrt{ (\Tr{L^\dag L}) (\Tr{K^\dag K}) } \leq \sqrt{ (1/2) \times (1/2) } = 1/2$ holds from the Cauchy-Schwarz inequality.
The corresponding term for the Choi operator of $\mathrm{id} \otimes \dcal$ satisfies $| \Tr \frac{I}{2} | = 1$,
and thus there are no concatenation $\bar{\ccal}_{\acal}^{K} \circ {\ccal}_{\bcal}^L$ that achieves $\mathrm{id} \otimes \dcal$.

The controlled quantum operation defined by Eq.~\eqref{eq:def_cc_2op} provides controlled quantum operation up to certain degree.
However, it does not cover the whole set of controlled quantum operations in general.
Note that if one of the quantum operation to be controlled is a unitary operation, the concatenation Eq.~\eqref{eq:def_cc_2op} provides the most general definition.
In particular, the general definition of controlled quantum operations leads to a non-trivial correlation between the two controlled quantum operations if interpreted as a concatenation of two controlled quantum operations.
For the example of $\mathrm{id} \otimes \dcal$, one possible Kraus representation is given by 
$\{ \frac{1}{2} ( \proj{0} \otimes \sigma_i + \proj{1} \otimes \sigma_i) \}_{i=0}^3$, 
where $\sigma_i$ denotes the Pauli operators as $ \sigma_0 = I, \sigma_1 = X, \sigma_2 = Y, \sigma_3 = Z $.
We can see that the same Kraus operator is applied regardless of the control qubit.
If we interpret as a concatenation of two controlled quantum operations, 
there is a non-trivial correlation between them which allows the same Kraus operator to be applied.

Note that while $\mathrm{id} \otimes \dcal$ provides a larger coherence in terms of the norm of the off-diagonal terms,
a larger coherence does not necessarily always provide useful applications.
As we also discuss in the next subsection, the concatenation of two controlled depolarizing channel, that is,
the one defined by Eq.~\eqref{eq:def_cc_2op} with $K = L = \frac{I}{2}$, provides a non-trivial effect identical to quantum switch as shown in Ref.~\cite{controlchannel1}.
Since the Kraus representation of such a controlled quantum operation is given by $\{ \frac{1}{2} ( \proj{0} \otimes I + \proj{1} \otimes I), \frac{1}{2}\proj{0} \otimes X, \frac{1}{2}\proj{0} \otimes Y, \frac{1}{2}\proj{0} \otimes Z, \frac{1}{2}\proj{1} \otimes X, \frac{1}{2}\proj{1} \otimes Y, \frac{1}{2}\proj{1} \otimes Z \}$,
only the Kraus operator $\frac{1}{2} ( \proj{0} \otimes I + \proj{1} \otimes I)$ contributes to the coherence in contrast to the case of $\mathrm{id} \otimes \mathcal{D}$ where all Kraus operators contributes to the coherence.

\subsection{Relationship between Controlled Quantum Operations and Quantum Switch}\label{sec:cc_qs}

In this subsection, we investigate the relationship between controlled quantum operations and quantum switch based on our definition of controlled quantum operations.
In Ref.~\cite{controlchannel1}, 
it is pointed out that the action of quantum switch on the depolarizing channels presented in Ref.~\cite{indefinite2} can be obtained by considering controlled depolarizing channels.
That is, by taking $K = L = \frac{1}{d} I$ in our definition for controlled quantum operations, 
the action of concatenation of two controlled depolarizing channels $\bar{\ccal}_{\mathcal D}^{L} \circ {\ccal}_{\mathcal D}^K = {\ccal}_{\mathcal D}^K \circ \bar{\ccal}_{\mathcal D}^{L}$ is equivalent to the action of quantum switch on depolarizing channels.
To simplify the problem, here we consider only two-dimensional case.
The Kraus operators for the depolarizing channel ${\mathcal D}$ is given by $\{ \frac{1}{2} \sigma_i \}_{i=0}^3$, where $\sigma_i$ denotes the Pauli operators.
The Choi operator of the output quantum operation of quantum switch is calculated as
\begin{align}
&\proj{00} \otimes \frac{I}{2} + \proj{11} \otimes \frac{I}{2} \notag \\
&\quad+ \ketbra{00}{11} \otimes \sum_{i,j} \frac{1}{2^4} \kketbra{\sigma_i \sigma_j}{\sigma_j \sigma_i} + h.c. \notag \\
&= \proj{00} \otimes \frac{I}{2} + \proj{11} \otimes \frac{I}{2} \notag \\
&\quad+ (\ketbra{00}{11} + \ketbra{11}{00}) \otimes \frac{1}{4} \pproj{I} \label{eq:cc_qs_action_qsdep}
\end{align}
by using the commutation relation of the Pauli operators.
The resulting quantum operation of concatenation of two controlled depolarizing channels, 
 $\bar{\ccal}_{\mathcal D}^{L} \circ {\ccal}_{\mathcal D}^K = {\ccal}_{\mathcal D}^K \circ \bar{\ccal}_{\mathcal D}^{L} $, is given by
\begin{align}
&\proj{00} \otimes \frac{I}{2} + \proj{11} \otimes \frac{I}{2} \notag \\
&\quad+ \ketbra{00}{11} \otimes \kketbra{L}{K} + \ketbra{11}{00} \otimes \kketbra{K}{L},
\end{align}
where $K$ and $L$ denotes the Kraus operators defining the controlled quantum operation.
It is easy to see that the two resulting quantum operations coincide if we take $K = L = \frac{1}{2} I$ as Ref.~\cite{controlchannel1} pointed out.

In Ref.~\cite{qs_comm_theory1,qs_comm_theory2}, 
the authors pointed out that while the input state passes through both depolarizing channels in the case of quantum switch,
it passes through only a single depolarizing channel in the controlled depolarizing channel case.
In fact, the actions of both cases coincide because concatenations of depolarizing channels are depolarizing channel.
Moreover, the authors pointed out that the concatenation of two controlled depolarizing channel is different from a single controlled depolarizing channel.
In our formalism, this fact is also obvious.
Assuming that the two depolarizing channels are characterized by Kraus operators $K, L$,
then the concatenated one is characterized by $LK$.
While $LK$ is a Kraus operator for the concatenated channel, such a product does not cover the whole set of Kraus operators of the concatenated channel.
For example, for the depolarizing channel case, while $\frac{1}{2} I$ is a Kraus operator of a single channel, it cannot be a product of two Kraus operators of the depolarizing channel,
i.e., of the form $LK$ with $K, L$ being the Kraus operators of the depolarizing channel.

In this subsection, we also show that if the input quantum operation is different from the depolarizing channel, such coincidence does not happen in general,
not only because the concatenation of two channels is not the same as the original one, but also the coherent term cannot be the same.
In particular, we consider the case where the input quantum operation ${\acal}$ is given by the Kraus operators $\{ \alpha_i \sigma_i \}_{i=0}^3$ with $ \alpha_i \geq 0$ satisfying $\sum \alpha_i^2 =1$.
Note that the depolarizing channel corresponds to $\alpha_i = 1/2$ for all $i$.
The action of quantum switch on this quantum channel is given by
\begin{align}
&\proj{00} \otimes J_{\acal \circ \acal} + \proj{11} \otimes J_{\acal \circ \acal} + \ketbra{00}{11} \otimes B + h.c.,
\end{align}
where the off-diagonal term $B$ is given by
\begin{align}
B &= \sum_i \alpha_i^4 \pproj{I} + 2[ (\alpha_0 \alpha_1)^2 - (\alpha_2 \alpha_3)^2 ] \pproj{X} \notag \\
&\quad+  2[ (\alpha_0 \alpha_2)^2 - (\alpha_1 \alpha_3)^2 ] \pproj{Y} \notag \\
&\quad+  2[ (\alpha_0 \alpha_3)^2 - (\alpha_1 \alpha_2)^2 ] \pproj{Z}. \label{eq:cc_qs_action_qs}
\end{align}
On the other hand, if we consider the concatenation of two controlled versions, which are characterized by $K,L$, respectively,
the resulting quantum operation $\bar{\ccal}_\acal^{L} \circ {\ccal}_\acal^K$ is given by
\begin{align}
\proj{00} \otimes J_\acal + \proj{11} \otimes J_\acal
+ \ketbra{00}{11} \otimes \kketbra{L}{K} + h.c.
\end{align}
Here the two operators $K,L$ have to satisfy
\begin{align}
K = \sum_{i} \beta_{i} (\alpha_i \sigma_i) \quad \sum_i \abs{\beta_i}^2 \leq 1, \\
L = \sum_{i} \gamma_{i} (\alpha_i \sigma_i) \quad \sum_i \abs{\gamma_i}^2 \leq 1,
\end{align}
and thus, we obtain
\begin{align}
\kketbra{L}{K} = \sum_{i,j} \gamma_{i} \beta_{j}^* \alpha_i \alpha_j \kketbra{\sigma_i}{\sigma_j}. \label{eq:cc_qs_action_cc}
\end{align}

From Eq.~\eqref{eq:cc_qs_action_qs}, we can see that unless $\alpha_i = 1/2$ for all $i$ or $\alpha_i = 1$ for one $i$ and otherwise $0$, which correspond to the depolarizing channel and the Pauli operations, respectively, 
at least one of $\pproj{\sigma_i}$ for $i \neq 0$ remains.
Assuming that $\pproj{\sigma_1}$ remains.
Then, in order to let the same term in Eq.~\eqref{eq:cc_qs_action_cc} be non-zero, it is required that $\alpha_1,\beta_1,\gamma_1 \neq 0$.
Also, the term $\pproj{I}$ in Eq.~\eqref{eq:cc_qs_action_qs} is non-zero, and it is required that $\alpha_0,\beta_0,\gamma_0 \neq 0$ from Eq.~\eqref{eq:cc_qs_action_cc}.
However, this indicates that the term $\kketbra{\sigma_0}{\sigma_1}$ is also non-zero in Eq.~\eqref{eq:cc_qs_action_cc}, 
where such term does not exist in Eq.~\eqref{eq:cc_qs_action_qs}.
Thus, the two resulting quantum operations can coincide only if the input operation is the depolarizing channel or the Pauli operations among the quantum operation given by the Kraus operators $\{ \alpha_i \sigma_i \}_{i=0}^3$.

If we consider the controlled version of the concatenation of two channels, $\bar{\ccal}_{\acal \circ \acal}^{L} \circ {\ccal}_{\acal \circ \acal}^K$,
the diagonal terms coincide, but the off-diagonal terms still cannot coincide.
The concatenation of two quantum channel $J_{\acal \circ \acal}$ is given by $\{ \alpha'_i \sigma_i \}_{i=0}^3$ where
\begin{gather}
\alpha'_0 = \sqrt{ \sum_i \alpha_i^4 } \\
\alpha'_1 = \sqrt{ 2 (\alpha_0 \alpha_1)^2 + 2 (\alpha_2 \alpha_3)^2 } \\
\alpha'_2 = \sqrt{ 2 (\alpha_0 \alpha_2)^2 + 2 (\alpha_1 \alpha_3)^2 } \\
\alpha'_3 = \sqrt{ 2 (\alpha_0 \alpha_3)^2 + 2 (\alpha_1 \alpha_2)^2 }.
\end{gather}
Here the two operators $K,L$ have to satisfy
\begin{align}
K = \sum_{i} \beta_{i} (\alpha'_i \sigma_i) \quad \sum_i \abs{\beta_i}^2 \leq 1, \\
L = \sum_{i} \gamma_{i} (\alpha'_i \sigma_i) \quad \sum_i \abs{\gamma_i}^2 \leq 1,
\end{align}
and thus, we obtain
\begin{align}
\kketbra{L}{K} = \sum_{i,j} \gamma_{i} \beta_{j}^* \alpha'_i \alpha'_j \kketbra{\sigma_i}{\sigma_j}.
\end{align}
Similarly, we can see that unless $\alpha_i = 1/2$ for all $i$ or $\alpha_i = 1$ for one $i$ and otherwise $0$, i.e., the depolarizing channel and the Pauli operations,
the two resulting quantum operations cannot coincide.
Note that here a range of the coherence between two quantum channels to be concatenated is allowed as we consider the controlled version of ${\acal} \circ {\acal}$, i.e., ${\ccal}_{\acal \circ \acal}^K$.
This also includes the case of the concatenation of two independently controlled channel, i.e., ${\ccal}_\acal^{K_2} \circ {\ccal}_\acal^{K_1}$,
because if $K_1$ and $K_2$ are the Kraus operators for the quantum operation ${\acal}$, then it is also possible to choose $K = K_2 K_1$ as a Kraus operator for ${\acal} \circ {\acal}$.
The inverse is not possible in general, and unless the coherent control of ${\mathcal D} \circ {\mathcal D}$ is allowed, the controlled depolarizing channel does not have the same action as the output operation of quantum switch,
that is, there exists no $K_1, K_2, L_1, L_2$ such that $\bar{\ccal}_{\mathcal D}^{L_2} \circ \bar{\ccal}_{\mathcal D}^{L_1} \circ {\ccal}_{\mathcal D}^{K_2} \circ {\ccal}_{\mathcal D}^{K_1}$ coincides with Eq.~\eqref{eq:cc_qs_action_qsdep}.

\section{Definition of the controlled quantum combs}\label{sec:controlled_quantum_combs}

A quantum operation transforms a given quantum state to another quantum state.  
Similarly, we can define a higher-order transformation, a transformation of a quantum operation to another quantum operation.
Quantum mechanically implementable transformations between quantum operations are investigated in Ref.~\cite{comb1,comb2}, and we summarize the relevant results for this paper in the following.

\begin{figure}
  \includegraphics[width=0.85\linewidth]{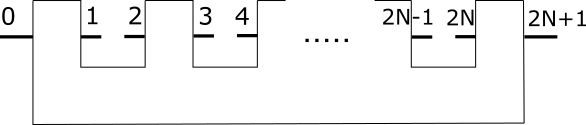}
\caption{A quantum comb with $N$-slot of the quantum operations.  
It is an abstract description of a quantum circuit which calls $N$ quantum operations in turn.  
The line with index $i$ represents the Hilbert space $\hcal_i$.  
The $k$-th quantum operation is the quantum operation which transforms a state on $\hcal_{2k-1}$ to that of $\hcal_{2k}$.  
The resulting quantum operation is a quantum operation transforming a state on $\hcal_0$ to $\hcal_{2N+1}$.}
\label{fig:quantum_comb}
\end{figure}

Mathematically, we consider the situation that we transform $N$ quantum operations ${\mathcal A}_k : L(\hcal_{2k-1}) \to L(\hcal_{2k})$ for $k=1, \ldots ,N$ to a target quantum operation ${\mathcal A}_0 : L(\hcal_{0}) \to L(\hcal_{2N+1})$.
Since any quantum operation can be described uniquely by its Choi operator, 
higher-order transformations between quantum operations can be described as transformations between the corresponding Choi operators.
We denote this transformation as ${\mathcal S} : L( \hcal_1 \otimes \hcal_2 \otimes \cdots \otimes \hcal_{2N} ) \to L( \hcal_0 \otimes \hcal_{2N+1})$,
\begin{align}
{\mathcal S} \left[ \bigotimes_k J_{{\mathcal A}_k}  \right] = J_{{\mathcal A}_0}.
\end{align}
The transformation ${\mathcal S}$ is linear, 
and can be described by an operator ${\mathcal J}_{\mathcal S} \in L( \hcal_0 \otimes \hcal_1 \otimes \cdots \otimes \hcal_{2N+1}) $, 
which is called as the Choi operator of the higher-order transformation ${\mathcal S}$.
In the quantum circuit formalism presented in \cite{comb1,comb2}, it is assumed that the quantum circuits implementing ${\mathcal A}_k$ can be used only once for each $k$ in turn.
Then the conditions for the transformation described by ${\mathcal J}_{\mathcal S}$ are given by
\begin{gather}
 {\mathcal J}_{\mathcal S} \geq 0 \label{eq:condition_comb_cp}\\
\Tr_{2k+1} {\mathcal J}_{\mathcal S}^{(2k+1)} = \Tr_{2k,2k+1} {\mathcal J}_{\mathcal S}^{(2k+1)} \otimes \frac{I_{2k}}{d_{2k}},
\label{eq:condition_seq}
\end{gather}
for $k=0,1,\ldots,N$, where ${\mathcal J}_{\mathcal S}^{(2k+1)} := \Tr_{2k+2,\ldots,2N+1} {\mathcal J}_{\mathcal S}$,
and $d_{2k}$ is the dimension of $\hcal_{2k}$.
This type of transformations of quantum operations to another quantum operation is called a {\it quantum comb} and it is represented as in the diagram in Fig.~\ref{fig:quantum_comb}.

Since a quantum comb is a linear and completely positive map similarly to quantum operations, 
${\mathcal S}$ can be represented by the Kraus operators $\{ S_i \}$ as follows.
\begin{align}
 {\mathcal S} \left[ J \right] = \sum_i S_i J S^\dagger_i,
\end{align}
where $S_i : \hcal_1 \otimes \hcal_2 \otimes \cdots \otimes \hcal_{2N} \rightarrow \hcal_0 \otimes \hcal_{2N+1}$.  
The Kraus representation and the Choi representation are related as
\begin{align}
 {\mathcal J}_{\mathcal S} = \sum_i \pproj{S_i}.
\end{align}

Note that the conditions for $S_i$ is not $\sum_i S_i^\dag S_i = I$, 
which would be the condition for a quantum operation to be trace-preserving.
Instead, the conditions for $S_i$ are determined by the conditions given by Eq.~\eqref{eq:condition_seq}.
In Appendix~\ref{ap:comb_kraus}, we rewrite this condition in terms of the Kraus representation.

Similarly to quantum operations, which can be implemented by a quantum circuit by adding an auxiliary system, any quantum comb can be implemented by a quantum circuit by adding an auxiliary system and inserting quantum gates before and after the input operations \cite{comb1,comb2}.   Note that the circuit implementation of a quantum comb is not unique, similarly to the case of a quantum operation.

In the following, we define the controlled version of a quantum comb analogous to the quantum operation case of Eq.~\eqref{eq:def_cck} shown in the previous section.
In the definition of a controlled quantum comb, it is not straightforward to define an identity comb corresponding to the identity operation required for defining controlled quantum operations.
In this paper, we consider the following quantum comb as the identity comb.
Assuming that $\dim \hcal_{2k} = \dim \hcal_{2k+1} $, we define the identity comb $\mathcal{S}_{\rm id}$, 
in which the state in $\hcal_{2k}$ is unchanged and transferred to $\hcal_{2k+1}$. This quantum comb is represented as
\begin{align}
 \mathcal{S}_{\rm id} [J] = S_{\rm id} J S_{\rm id}^\dagger, 
\end{align}
with the Kraus operator given by
\begin{align}
 S_{\rm id} = \left( \bigotimes_{k=0}^{N} \bbra{I}_{2k, 2k+1} \right) \kket{I}_{0,0} \kket{I}_{2N+1, 2N+1}.
\end{align}
The action of this quantum comb is given by
\begin{align}
 {\mathcal S}_{\rm id} \left[ \bigotimes_{k=1}^N J_{{\mathcal A}_k} \right] = J_{{\mathcal A}_N \circ \cdots \circ {\mathcal A}_1}. 
\end{align} 
Note that  the following arguments of this section can be generalized to the case that the quantum comb is described by a single Kraus operator, instead of this identity comb.

\begin{figure}
  \includegraphics[width=0.85\linewidth]{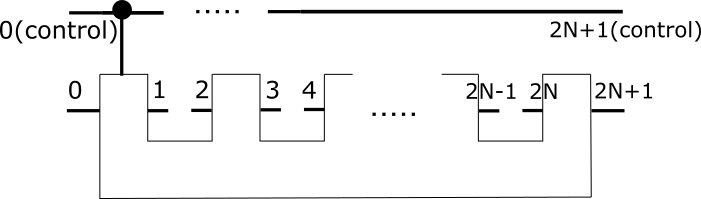}
\caption{A controlled quantum comb.  
The state of the control qubit belongs to the initial and the final state.} \label{cc4}
\end{figure}

Analogous to the controlled quantum operation defined by Eq.~\eqref{eq:def_cck}, 
we define the controlled version of quantum comb ${\mathcal C}_{\mathcal S}$ as in Fig.~\ref{cc4} by the following Choi  operator
\begin{align}
 {\mathcal J}_{ {\mathcal C}_{\mathcal S}^{ S } } &:=  \proj{00} \otimes {\mathcal J}_{{\mathcal S}_{\rm id}} + \proj{11} \otimes \mathcal{J}_{\mathcal S} \nonumber \\
 &\quad + \ketbra{00}{11} \otimes \kketbra{I}{S} + \ketbra{11}{00} \otimes \kketbra{S}{I},  \label{eq:def_cck_comb}
\end{align} 
where $S = \sum_i \alpha_i^* S_i$ with $\sum_i \abs{\alpha_i}^2 = 1$ and ${\mathcal S}_{\rm id}$ is the identity comb.  
Notice that if we trace out the final system, which includes the control qubit system, the third and fourth terms vanishes.
Thus, it is clear if the quantum comb to be controlled satisfies the condition given by Eq.~\eqref{eq:condition_seq},
the controlled version also satisfies the same condition.

Moreover, as in the quantum operation case, we can define the {\it most coherently controlled quantum comb} in terms of the operator $S$ by
\begin{align}
 S = \sum_{i\in B_{\rm max} } \alpha_i^\ast \tilde{S_i}, \blank \sum_i \abs{\alpha_i}^2 = 1, 
\end{align}
where $\{ \tilde{S_i} \}$ is an orthogonal Kraus representation of the quantum comb ${\mathcal S}$ and 
\begin{align}
 B_{\rm max}:= \left\{ i~ \Big\vert~ \forall j, ~ \tr{\tilde{S_i}^\dagger \tilde{S_i}} \geq \tr{\tilde{S_j}^\dagger \tilde{S_j}} \right\}.
\end{align}
The most coherently controlled quantum combs are expected to provide larger coherence in the resulting controlled quantum operations, 
and are as important as the most coherent controlled quantum operations.
In particular, the most coherently controlled quantum combs of certain types can be used to produce the most and fully coherently controlled quantum operations as we show in the next section.

\subsection{Controlled quantum combs switching between two quantum combs}

In the quantum comb case, it is also possible to consider a controlled quantum comb that switches between two quantum combs,
namely, a controlled quantum comb which applies quantum comb $\mathcal{S}$ if the control qubit is in $\ket{0}$,
and applies quantum comb $\mathcal{T}$ if the control qubit is in $\ket{1}$.
The quantum comb given by the Choi operator
\begin{align}
&\proj{00} \otimes  {\mathcal J}_{{\mathcal S}} + \proj{11} \otimes  {\mathcal J}_{{\mathcal T}} \notag \\
&\quad+ \ketbra{00}{11} \otimes \kketbra{S}{T} + \ketbra{11}{00} \otimes \kketbra{T}{S}, \label{eq:def_ccomb_2qcomb}
\end{align}
provides such a controlled quantum comb,
where $S$ and $T$ denotes the Kraus operators defining each controlled quantum comb ${\ccal}_{\mathcal S}^{S}$ and ${\ccal}_{\mathcal T}^{T}$.
In the quantum comb case, the controlled quantum comb defined by Eq.~\eqref{eq:def_ccomb_2qcomb} cannot be simply implemented by a concatenation of each controlled quantum comb ${\ccal}_{\mathcal S}^{S}$ and ${\ccal}_{\mathcal T}^{T}$, 
as there is an order between input quantum operations and the quantum comb.
However, Eq.~\eqref{eq:def_ccomb_2qcomb} define a valid controlled quantum comb because it satisfies the condition for the quantum comb,
and it preserves the coherence up to certain degree.

\section{Controllization of unitary operations with a controlled neutralization comb}

\subsection{Neutralization comb and controlled quantum operations}

We investigate the relationship between controlled quantum operations and controlled quantum combs defined in the previous sections to seek applications of controlled quantum combs in quantum computation.
We consider a class of quantum combs which we call \emph{neutralization combs},
i.e., quantum combs transforming any input quantum operation into the identity operation.
A quantum comb $\ncal$ which takes $N$ quantum operations ${\mathcal A}_1, \ldots, {\mathcal A}_N$ as inputs is a neutralization comb 
for a set of input operations $S_N$ if 
\begin{align}
 {\mathcal S} \left[ \bigotimes_k J_{{\mathcal A}_k} \right] = J_{\rm id}, \label{defneutralizationcomb}
\end{align} 
holds for all $({\mathcal A}_1, \ldots, {\mathcal A}_N) \in S_N$,
where $J_{\rm id}$ is the Choi operator of the identity operation.  %
Note that the condition given by Eq.~(\ref{defneutralizationcomb}) does {\it not} uniquely determine a neutralization comb, since there are many quantum combs satisfying Eq.~(\ref{defneutralizationcomb}) forming a class of neutralization combs.

When we have quantum operations ${\mathcal A}_1, \ldots, {\mathcal A}_N$ as input operations of a controlled neutralization comb,
the resulting quantum operation is a controlled quantum operation of ${\mathcal A}_N \circ \cdots \circ {\mathcal A}_1$.
That is, if the control qubit is in $\ket{0}$, the controlled quantum operation applies the identity operation, 
and if the control qubit is in $\ket{1}$, it applies ${\mathcal A}_N \circ \cdots \circ {\mathcal A}_1$. (See Axiom~\ref{axm:cc})
From now on, for adapting the standard notation of controlled quantum operations, we exchange the state of the control qubit when the identity comb is applied and when a neutralization comb is applied.
Namely, we apply the neutralization comb if the control qubit is in $\ket{0}$ and apply the identity comb if the control qubit is in $\ket{1}$, so that the role of the control qubit of the resulting controlled quantum operation coincides with {the standard} definition of controlled quantum operations.

\begin{figure}
  \includegraphics[width=0.6\linewidth]{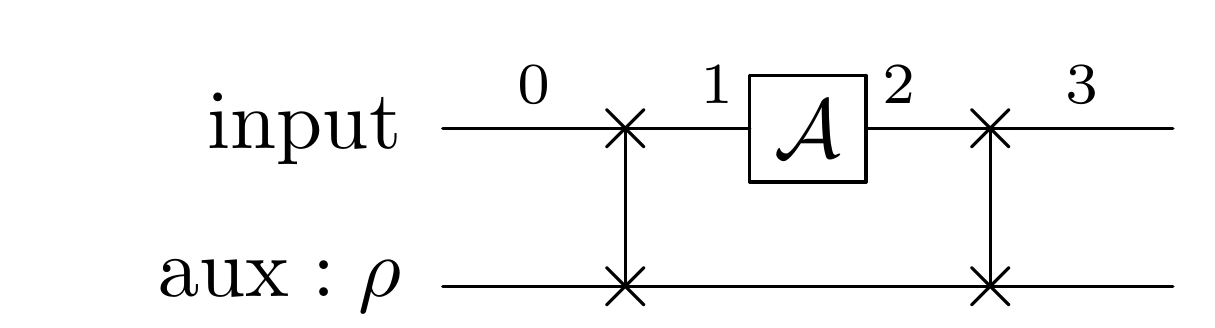}
\caption{A neutralization comb with a single input quantum operation ${\mathcal A}$ defined by Eq.~\eqref{eq:prepare_and_traceout} with $N=1$.
Any input state of $\hcal_0$  is sent to the output state without any change regardless of the quantum operation ${\mathcal A}$.} \label{fig:neutralization_comb_1}
\end{figure}

One way to implement a neutralization comb is to apply the input quantum operations on the auxiliary system, and then discarding the auxiliary system. 
Mathematically, this neutralization comb is described as
\begin{align}
{\mathcal J}_{\ncal} = \pproj{I}_{0, {2N+1}} \otimes \rho_{\hcal_{in} } \otimes I_{\hcal_{out}}, \label{eq:prepare_and_traceout}
\end{align}
where $\hcal_{in} = \bigotimes_{k=1}^{N} \hcal_{2k-1}, \hcal_{out} = \bigotimes_{k=1}^{N} \hcal_{2k}$ and $\rho \in L(\hcal_{in})$ is a quantum state that is initially prepared in the auxiliary system.
The corresponding quantum circuit of this neutralization comb for $N=1$ is shown in Fig.~\ref{fig:neutralization_comb_1}.

The first guess is to simply use this neutralizing comb for defining a most coherently controlled neutralization comb and then obtain the most coherently controlled quantum operation. 
However, the most coherently controlled neutralization comb does not necessarily provide the most coherently controlled quantum operation in general.
For example, consider the neutralization comb given by Eq.~\eqref{eq:prepare_and_traceout} for $N=1$.
When $\mathcal{A}_1$ is a single unitary operation described by $U$, the corresponding most coherently controlled operation is given by the controlled unitary operation ${\mathcal C}_U$ defined as Eq.~\eqref{eq:cont_u}.
However, it is shown that the controlled unitary operation is not implementable in this situation~\cite{controllization1,controllization2,controllization3,controllization4}, regardless of how the controlled neutralization comb is defined.

\subsection{Neutralization comb for unitary operations with a known eigenstate}\label{sec:known_eigenstate}

\begin{figure}
\includegraphics[width=0.6\hsize]{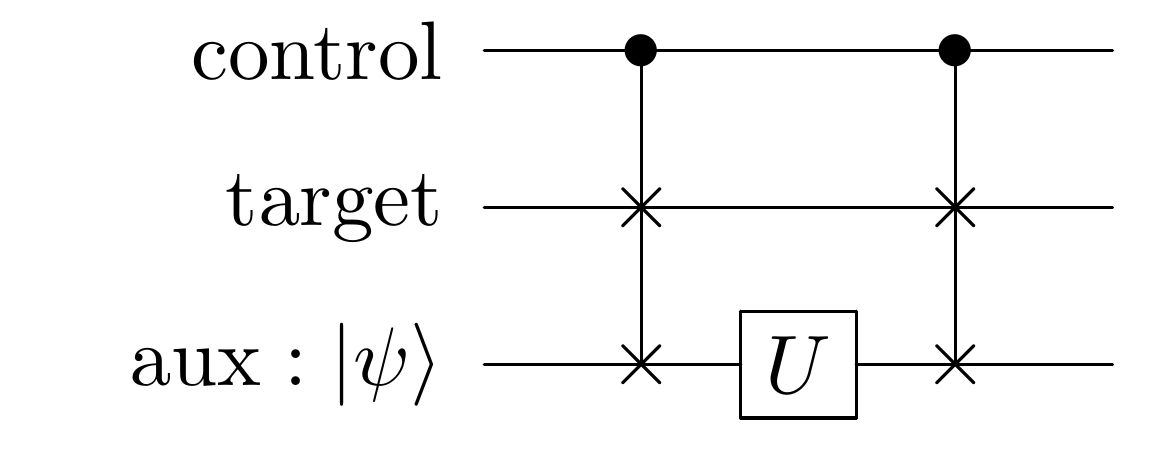}
\caption{Quantum circuit for the controlled neutralization defined by Eq.~\eqref{eq:controlled_neutralization_1}.
If the unitary operator $U$ has an eigenstate of $\ket{\psi}$, 
this quantum circuit exactly implements the corresponding controlled unitary operation.
}\label{fig:controlled_neutralization_1}
\end{figure}

Nevertheless, the most coherently controlled neutralization comb can implement the action of the most coherently controlled quantum operation  by restricting the set of the input quantum operations.
One example of a restricted set that the most coherently controlled quantum operation can be implemented
is the set of unitary operations of which one of the eigenstate of the unitary operator $U$ is given, namely,
$S_N = \{ U \mid U \ket{\psi} = e^{i \theta_U} \ket{\psi}, U \in \mathrm{U}(d) \}$ where $\ket{\psi}$ is an eigenstate and $\theta_U$ is an arbitrary phase.
Consider the controlled neutralization comb given by Eq.~\eqref{eq:prepare_and_traceout}.
It is easy to see that if we set the auxiliary state to be $\rho = \proj{\psi}$, the controlled unitary operation is implemented.
Mathematically, this neutralization comb is described by ${\mathcal J}_{\ncal} = \kketbra{I}{I}_{03} \otimes \proj{\psi}_{1} \otimes I_{2}$.
As shown in the previous sections, only the eigenvector which has the maximal norm contributes for the most coherently controlled comb.
In this case, it is possible to choose any element as $\kket{S_0} = \kket{I}_{03} \otimes \ket{\psi}_{1} \otimes \ket{\phi}_{2}$ with an arbitrary state $\ket{\phi}$. 
By requiring the controlled version of the identity operation $\mathrm{id}_{\hcal \rightarrow \kcal} $  is still the identity operation  $\mathrm{id}_{\hcal_C \otimes \hcal \rightarrow \kcal_C \otimes \kcal}$,
we obtain $\ket{\phi} = \ket{\psi^*}$, and the corresponding fully coherently controlled neutralization comb is given by
\begin{align}
 {\mathcal J}_{{\mathcal C}_\ncal} &=  \proj{00} \otimes \jcal_\ncal + \proj{11} \otimes \pproj{I} \notag \nonumber \\
 &+ \ketbra{00}{11} \otimes \kketbra{S_0}{I} + \ketbra{11}{00} \otimes \kketbra{I}{S_0}, \label{eq:controlled_neutralization_1}\\
 \kket{S_0} &= \kket{I}_{03} \otimes \ket{\psi}_1 \otimes \ket{\psi^*}_2.
 \end{align}
 A quantum circuit for this implementation of the neutralization comb is shown in Fig.~\ref{fig:controlled_neutralization_1}.
The action of this controlled neutralization comb $\jcal_{\ccal_\ncal}$ for $U$ is given as 
\begin{align}
&\Tr_{\hcal_{in}\hcal_{out}} [ \jcal_{\ccal_\ncal} ( \pproj{U}_{12} )^T ] \nonumber \\
&= \ketbra{00}{00} \otimes J_I
+ \ketbra{11}{11} \otimes J_U \nonumber \\
&\quad + \ketbra{11}{00} \otimes \bbra{U^*}_{12} ( \kketbra{I}{S_0}_{0123}) \kket{U^*}_{12} + h.c. \nonumber \\
&= \ketbra{00}{00} \otimes J_I
+ \ketbra{11}{11} \otimes J_U \nonumber \\
&\quad +  \ketbra{11}{00} \otimes \kketbra{e^{-i \theta_U} U}{I}_{03} + h.c. %
\end{align}
where the last equality holds due to
\begin{align}
\bbra{U^*}_{12} \kket{S_0}_{0123} &= \kket{I}_{03} \sum_i \bra{ii} (I \otimes U^T) \ket{\psi \psi^*} \nonumber \\
&= \kket{I}_{03} \sum_i \bra{ii} (U \otimes I) \ket{\psi \psi^*} \nonumber \\
&= \kket{I}_{03} e^{i \theta_U}.
\end{align}

\subsection{Controllization of divisible unitary operations}

If a unitary operation is generated by a time-independent Hamiltonian dynamics $U = e^{-i H t}$, 
division of the time evolution is possible, and by selecting the duration time to be $t/n$, it is possible to implement $U^{1/n} =  e^{-i H t/n}$.
For such divisible unitary operations, we can consider the controllization of $U$ by using $V = U^{1/n}$ for $n$ times.
In this subsection, we present two quantum algorithms for universal controllization of divisible unitary operations by utilizing the most coherently controlled neutralization comb.
The calculations are given in Appendix~\ref{ap:multicopy} and~\ref{ap:pauli_randomization}.
Here we only present the obtained quantum algorithms in terms of quantum circuits.

The first quantum algorithm utilizes $U^{1/d}$ for $d$ times with $d = \dim U$ to implement the desired controlled unitary operation $\ccal_U$ in an exact manner.
The corresponding quantum circuit is shown in Fig.~\ref{fig:neutralization_comb_2f},
and the calculations are presented in Appendix~\ref{ap:multicopy}.
Here we utilize the $d$-dimensional totally antisymmetric state $\ket{A_d}$ defined by
\begin{align}
\ket{A_d} = \frac{1}{\sqrt{d!}} \sum_{\sigma \in \mathcal{S}_d} 
\mathrm{sgn}(\sigma) \ket{\sigma(1)} \ket{\sigma(2)} \cdots \ket{\sigma(d)},
\end{align}
where $\mathcal{S}_d$ is the $d$-dimensional symmetric group and $\sigma$ denotes a permutation.

This algorithm achieves exact controllization because the $d$-dimensional totally antisymmetric state $\ket{A_d}$ satisfies $U^{\otimes d} \ket{A_d} = (\det{U}) \ket{A_d}$ for all $U \in \mathrm{U}(d)$,
and it allows a coherent deletion of the input unitary operations, i.e., the $d$ copies of $U$.
It is also possible to regard this algorithm as a modified version of Fig.~\ref{fig:controlled_neutralization_1},
where the auxiliary system is an eigenstate of the unitary operation $U^{\otimes d}$.

\begin{figure}
\includegraphics[width=0.95\hsize]{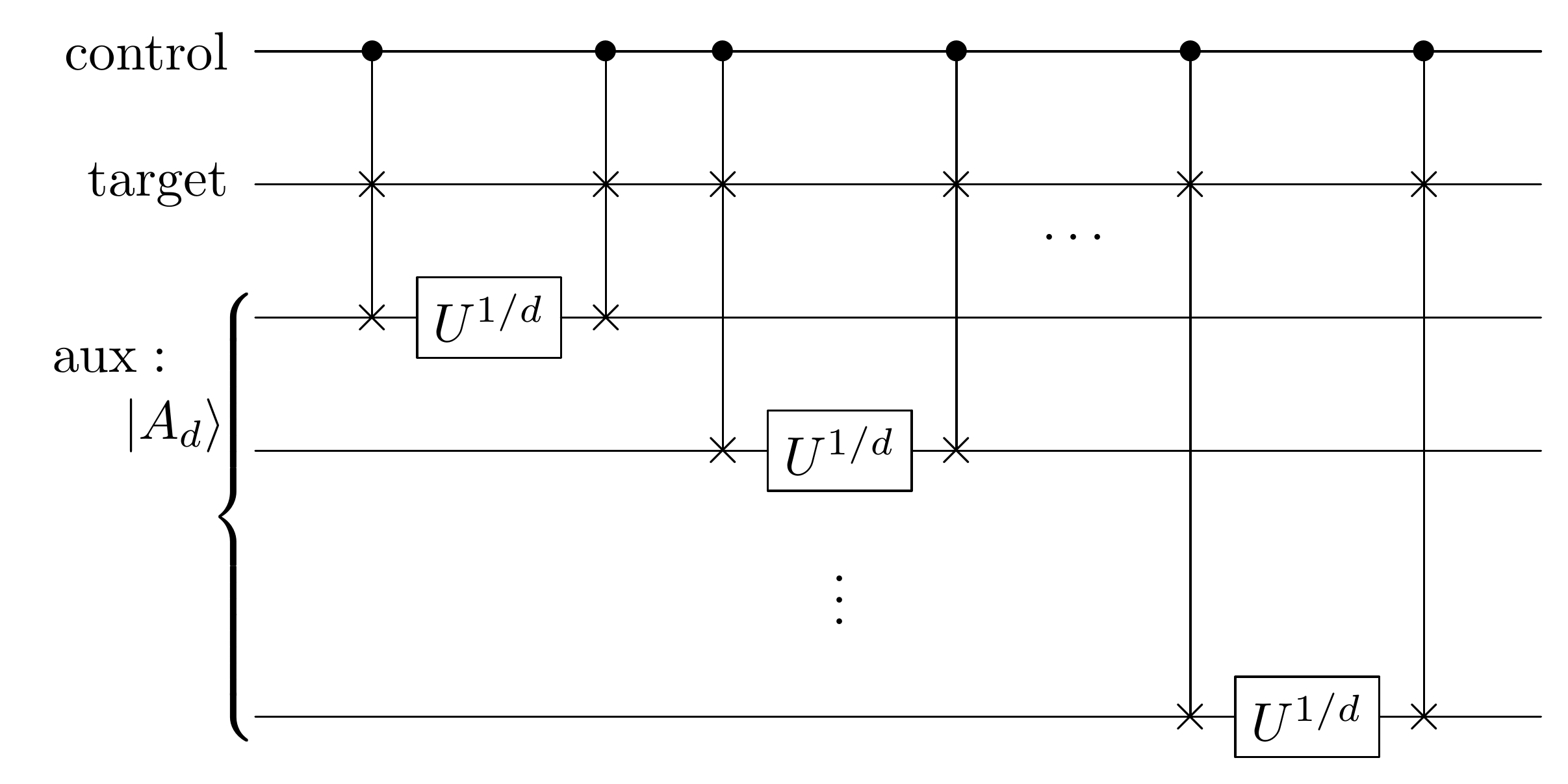}
\caption{Quantum circuit that exactly implements controlled unitary operation $\ccal_U$ with $d$ uses of $U^{1/d}$, where $d = \dim U$.
} \label{fig:neutralization_comb_2f}
\end{figure}

\begin{figure}[b]
\includegraphics[width=0.75\hsize]{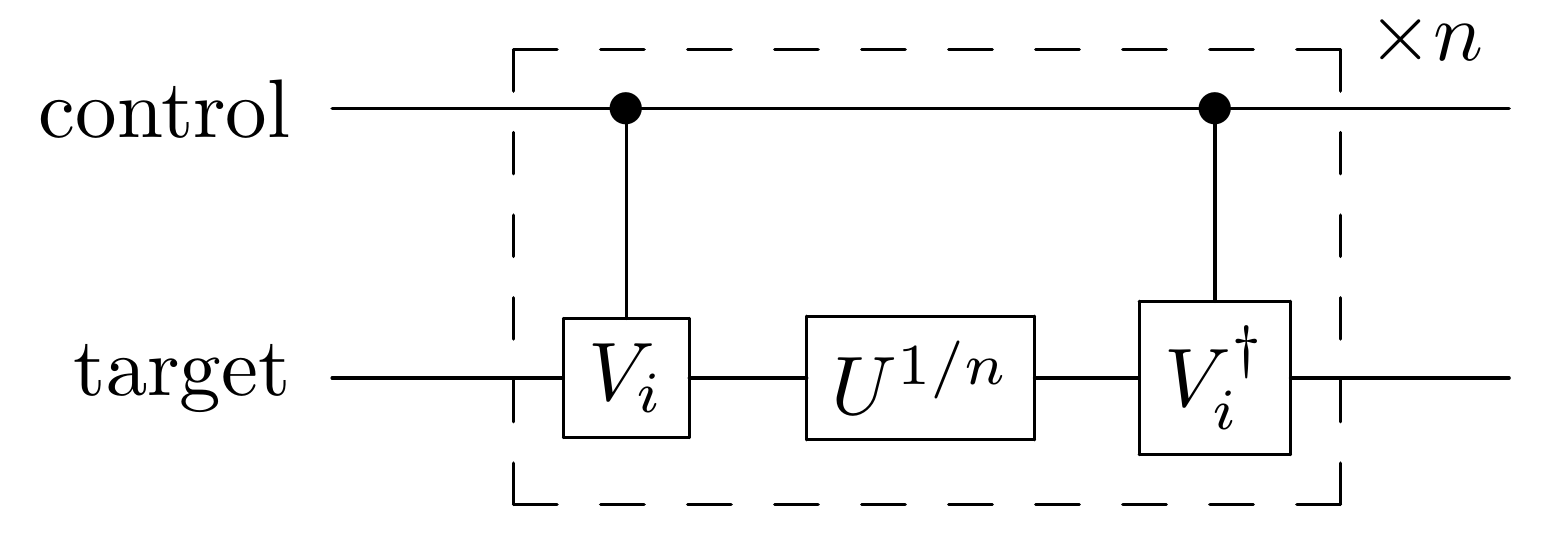}
\caption{Quantum circuit that approximately implements controlled unitary operation $\ccal_U$ by using $U^{1/n}$ for $n$ times with an error of $O(1/n)$.
The dotted box indicates that the quantum circuit inside the box is repeated for $n$ times.
The unitary operations $\{ V_i \}$ are chosen randomly from the Pauli operations $\{ I, X, Y, Z \}$ with an equal probability in each repetition.
}
\label{fig:controlled_randomization_f}
\end{figure}

The second quantum algorithm utilizes $U^{1/n}$ for $n$ times to implement the desired controlled unitary operation $\ccal_U$ for two-dimensional unitary operation in an approximate manner.
The corresponding quantum circuit is shown in Fig.~\ref{fig:controlled_randomization_f},
and the calculations are shown in Appendix~\ref{ap:pauli_randomization}.
In Ref.~\cite{pme}, a similar algorithm for implementing controllization was presented,
where the randomization is applied to an auxiliary system and its initial state is prepared in the maximally mixed state.
Our algorithm performs a randomization on the input unitary operation directly instead of on the auxiliary state as in the one presented in Ref.~\cite{pme}.
Thus, our algorithm can be implemented without using an auxiliary system.

Here the randomization is applied based on the Pauli operations $\{ V_i \} = \{ I, X, Y, Z \}$,
but it is also possible to consider a randomization based on the Clifford operators.
We compare both performance of the randomization by Pauli operations and Clifford operations in Appendix~\ref{ap:pauli_randomization} and~\ref{ap:clifford_randomization},
and we show that the randomization by Pauli operations has a better performance.
This is because the algorithm performs a randomization on the input operations but not states,
and the randomization by the Clifford operations randomizes the input operation ``too much''.

\section{Conclusion}

We have defined  a controlled quantum operation of a general deterministic quantum operation based on two physical implementations and a set of axioms, which coincide with each other.
We then  analyzed the coherence between the quantum operations on different control qubit states,
and gave a characterization of the controlled quantum operations that maximize the coherence,
which we call as  the most coherently controlled quantum operation.
This definition on quantum operation is extended  to quantum combs, 
and we defined controlled quantum combs and  the most coherently controlled quantum combs.
The definition of  the most coherently controlled quantum operation or comb can be generalized to the controlled  two arbitrary quantum operations or combs,
if one of them has a Kraus representation  consisting of a single Kraus operator.
We also discussed about the controlled two general quantum operations or combs from the previous definition,
but this method does not provide the most general controlled ones.
Thus, it is an open question how to define the most general controlled  two general quantum operations or combs,
and find the most coherent ones or the most optimized ones for certain tasks, e.g. preserving information, among them.
We note that it is also possible to consider a generalization to the case where the control system is a $d$-dimensional qudit, instead of a qubit, 
and applying different quantum operations or combs depending on the state of the control system.
 
We showed a relation between controlled quantum operations and controlled quantum combs, by introducing  the neutralization combs.
While the most coherently controlled quantum comb does not always implement the most coherently controlled quantum  operation in general,
we showed that this is possible by restricting the input quantum operations to be the same unitary  operation, and show that if the unitary operation $\ucal$ represented by a unitary operator $U$ is divisible to $d$ products of $U^{1/d}$, 
the most coherently controlled neutralization comb provide an implementation of the fully coherently controlled unitary operation of $\ucal$.
We also introduce approximate neutralization combs for Hamiltonian dynamics represented by a unitary operator $U=e^{-i H t}$, implemented by the basis randomization combs with the Pauli operators and the Clifford operators,
and show that the most coherently controlled basis randomization comb can be used for controllization of the Hamiltonian dynamics, which is an infinitely  divisible unitary operation.

Controllization of unitary operations have been considered in various previous works,
and we provided two new methods for implementing controllization of divisible unitary operations.
In particular, they present advantages compared to the previous works in that the first one can be implemented in an exact manner, 
and the second one requires no auxiliary system.
On the other hand, there are also a few results on avoiding the requirement of controllization in certain quantum algorithms.
For example, while DQC1 is an algorithm that originally utilizes a controlled unitary operation,
a modified version of DQC1 is proposed in Ref.~\cite{controllization1} which provides the same results without using a controlled unitary operation.
Not all quantum algorithms can avoid the usage of controlled unitary operations, 
and it is not known how to analyze the effect of the coherence of controlled unitary operations in a quantum algorithm utilizing controlled unitary operations.
We hope our framework sheds a new light in analyzing the coherence of controlled operations in quantum computing.

\section*{Acknowledgements}

This work was supported by MEXT Quantum Leap Flagship Program (MEXT Q-LEAP) Grant Number JPMXS0118069605 and JPMXS0120351339, Japan Society for the Promotion of Science (JSPS) by KAKENHI grant No.~17H01694, 18H04286, 18K13467 and Advanced Leading Graduate Course for Photon Science (ALPS).

\bibliographystyle{utphys}
\bibliography{refs}

\appendix

\section{Neutralization comb for the same input unitary operations}\label{ap:multicopy}

When multiple uses of the same unitary operation are available,
we show that another construction of the most coherently controlled unitary operation is possible. 
In this case, we assume that the input unitary operation is described by a $d$-dimensional unitary operator $U$,
and we use this unitary operation $n$ times. 
The set of input operations to be neutralized is given by $S_N = \{ (U_1, \ldots, U_d) \mid U_1 = \cdots = U_d = U \in \mathrm{U}(d) \}$.
Note that in this case, the output of the identity comb is given by $U^n$,
and the controlled quantum operation that we are aiming to implement is given by $\proj{0} \otimes I + \proj{1} \otimes  U^n $.

Similarly to the case of the single input operation analyzed in Sec.~\ref{sec:known_eigenstate}, we assume that the  neutralization comb is achieved by a preparation of an auxiliary state and then trace out the auxiliary system.
This neutralization comb can be written as
\begin{align}
{\mathcal J}_{\ncal} = \pproj{I}_{0, {2N+1}} \otimes \rho_{\hcal_{in} } \otimes I_{\hcal_{out}}. \label{eq:comb_prepare_traceout}
\end{align}
The controlled version of this neutralization comb is described by 
$\kket{S_0} = \lambda \kket{I} \otimes \ket{\psi} \otimes \ket{\phi}$ with arbitrary states $\ket{\psi}$, $\ket{\phi}$ and a normalization constant $| \lambda | \leq 1$.
The action of this controlled comb is
\begin{align}
&\Tr_{\hcal_{in}\hcal_{out}} [ \jcal_{\ccal_\ncal} ( \pproj{U}^{\otimes n}  )^T ] \nonumber \\
&= \ketbra{00}{00} \otimes J_I
+ \ketbra{11}{11} \otimes J_{U^n} \nonumber \\
&\quad + \ketbra{11}{00} \otimes \kket{U^n} \bbra{S_0} (\kket{(U^*)^n}) \nonumber \\
&\quad + \ketbra{00}{11} \otimes (\bbra{(U^*)^n}) \kket{S_0} \bbra{ U^n }.
\end{align}
As we require this to be the Choi operator of the controlled unitary operation,
\begin{align}
&\ketbra{00}{00} \otimes J_I + \ketbra{11}{11} \otimes J_{U^n} \notag \nonumber \\
& + \ketbra{11}{00} \otimes \kketbra{U^n}{I} + \ketbra{00}{11} \otimes \kketbra{I}{U^n}, \label{eq:cont_Un}
\end{align}
we obtain the condition for the off-diagonal coherence term
\begin{align}
(\bbra{(U^*)^n}) \kket{S_0} = \kket{I},
\end{align}
or equivalently,
\begin{align}
\lambda | \bbra{I} ( U^{\otimes n} \otimes I_\kcal )  (\ket{\psi} \otimes \ket{\phi}) | = 1.
\end{align}

Notice that the maximally entangled state can be written as $\kket{I} = \sum_i \ket{ii} = \sum_i \ket{\psi_i \psi_i^*}$, 
where $ \{ \ket{\psi_i} \} $ is an arbitrary basis,
the off-diagonal coherence term can be evaluated as
\begin{align}
\lambda \bbra{I} ( U^{\otimes n} \otimes I_\kcal )  (\ket{\psi} \otimes \ket{\phi}) = \lambda \bra{\phi^*} (U^{\otimes n}) \ket{\psi}
\end{align}
and its absolute value is $|\lambda| | \bra{\phi^*} (U^{\otimes n}) \ket{\psi} | $.
This can achieve 1 only if $|\lambda| = 1$ and $\ket{\psi} = e^{i\theta_U} U^{\otimes n} \ket{\phi}$.
Thus, we obtain the necessary condition that $\ket{\psi}$ is invariant under the action of $U^{\otimes n}$.
This condition is equivalent to the existence of a one-dimensional invariant subspace of $U^{\otimes n}$,
which, by considering the Schur-Weyl duality, happens if and only if $n$ is a multiple of $d = \dim U$.

\begin{figure}
  \includegraphics[width=0.9\linewidth]{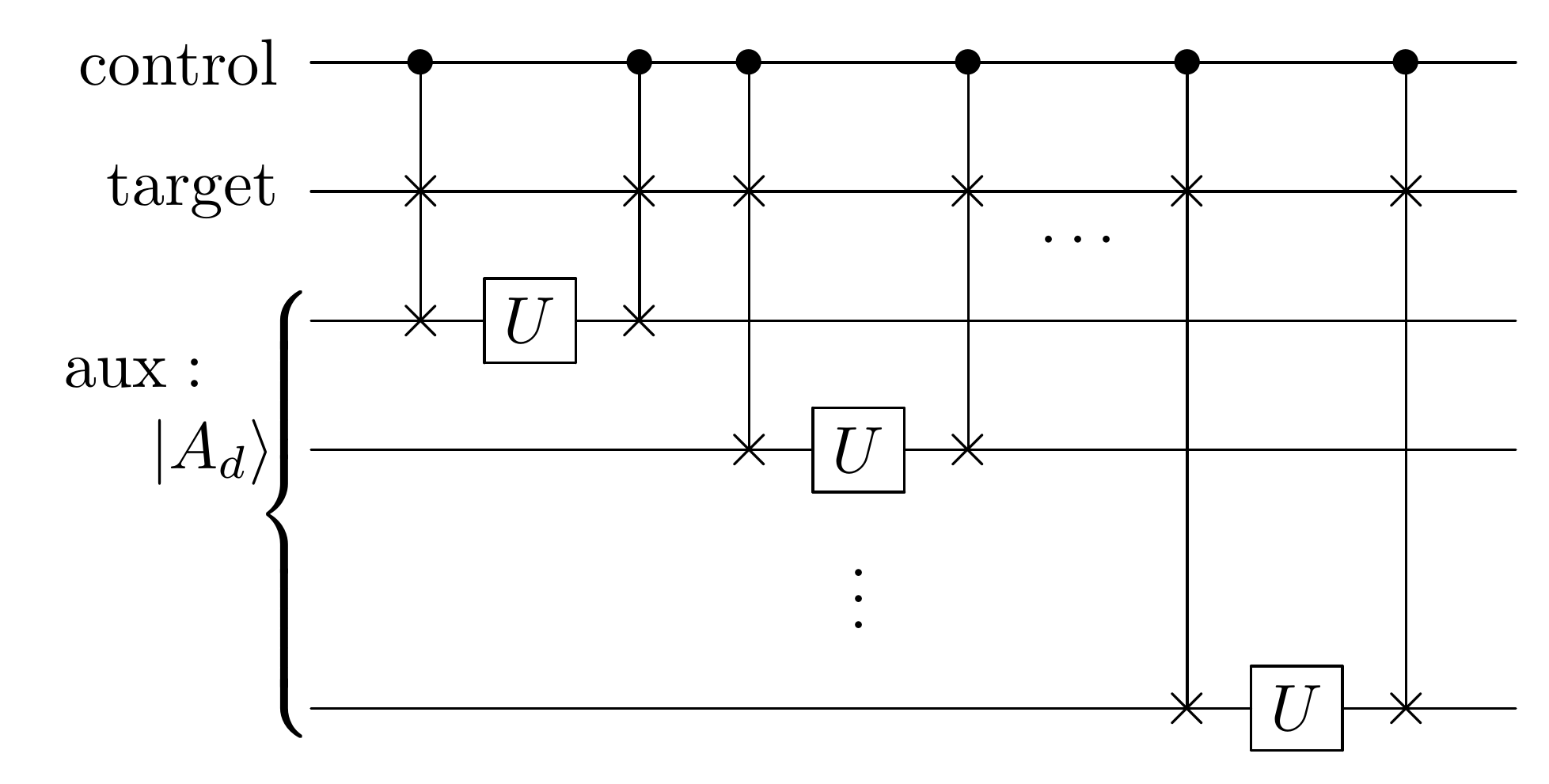}
\caption{Quantum circuit for the most coherently controlled neutralization comb with multiple uses of the same unitary operations $U$ defined by Eq.~\eqref{eq:def_controlled_comb_anti}.
This quantum circuit exactly implements the controlled unitary operation of $U^d$.
$\ket{A_d}$ denotes a totally antisymmetric state of a system with $\mathcal{H}=(\mathbb{C}^{d})^{\otimes d}$ satisfying $U^{\otimes d} \ket{A_d} = \det(U) \ket{A_d}$.
} \label{fig:neutralization_comb_2}
\end{figure}

Thus, the necessary condition for most (and fully) coherently controlled unitary operation Eq.~\eqref{eq:cont_Un} to be  implementable, i.e., $n$ is a multiple of $d$, is shown.
This condition is also the sufficient condition.
That is, if there exists an invariant state $\ket{\psi}$ under the action of $U^{\otimes n}$, 
the quantum circuit shown in Fig.~\ref{fig:neutralization_comb_2} implements the (fully coherently) controlled unitary operation.

If a unitary operation is generated by a time-independent Hamiltonian dynamics $U = e^{-i H t}$, division of the time evolution by $d$ is possible by selecting the duration time to be $t/d$, namely, $U^{1/d} =  e^{-i H t/d}$.
For such divisible unitary operations, we can consider the controllization of $U$ by using $V = U^{1/d}$ for $d$ times.   The quantum circuit shown in Fig.~\ref{fig:neutralization_comb_2} implements the desired controlled unitary operation $\ccal_U$.
In this case, if we take $U \in \mathrm{SU}(d)$, the ambiguity of the global phase $e^{2 \pi i /d}$ disappears as the global phase is multiplied $d$ times.

For completeness, the Choi operator for the most coherently controlled neutralization comb is given by
\begin{align}
 {\mathcal J}_{{\mathcal C}_\ncal} &=  \proj{00} \otimes \jcal_\ncal + \proj{11} \otimes \pproj{I} \notag \\
 &+ \ketbra{00}{11} \otimes \kketbra{S_0}{I} + \ketbra{11}{00} \otimes \kketbra{I}{S_0}, \label{eq:def_controlled_comb_anti}\\
 \kket{S_0} &= \kket{I}_{0,2N+1} \otimes \ket{A_d}_{\hcal_{in}} \otimes \ket{A_d}_{\hcal_{out}} .
\end{align}
Here $\ket{A_d}$ is the invariant state satisfying $U^{\otimes d} \ket{A_d} = (\det{U}) \ket{A_d}$ for all $U \in \mathrm{U}(d)$.
More explicitly, $\ket{A_d}$ is the $d$-dimensional totally antisymmetric state,
\begin{align}
\ket{A_d} = \frac{1}{\sqrt{d!}} \sum_{\sigma \in \mathcal{S}_d} 
\mathrm{sgn}(\sigma) \ket{\sigma(1)} \ket{\sigma(2)} \cdots \ket{\sigma(d)},
\end{align}
where $\mathcal{S}_d$ is the $d$-dimensional symmetric group and $\sigma$ denotes a permutation.
An iterative algorithm to generate $\ket{A_d}$ is shown in Ref.~\cite{antisymmetric}.

Remark that in this section, 
we assume that the Choi operator of the neutralization comb has the form of Eq.~\eqref{eq:comb_prepare_traceout},
which is implemented by first preparing a quantum state on the auxiliary system, and discard the auxiliary system at the end.
If we further restrict the initial state of the auxiliary state to be a pure state, 
the necessity of the requirement for the initial auxiliary state to be a one-dimensional invariant state is trivial since it is equivalent to an invariant pure state.
However, if we allow to prepare a mixed state for the initial state of the auxiliary system, the maximally mixed state, $I/d$, is invariant under the action of unitary operations. 
Although the invariant states exist both in the pure and mixed state, only the pure invariant state can contribute for exactly implementing the fully coherently controlled divisible unitary operation.   However, in approximate cases,  the maximally mixed state has been utilized for implementing controlled divisible unitary operation with a randomization algorithm shown in~\cite{pme}.

\section{Basis randomization comb with Pauli operators}\label{ap:pauli_randomization}

In this appendix and the next appendix, 
we consider an approximate neutralization comb employing random unitary operators, which we call a basis randomization comb.
The idea of using random unitary operators to implement controllization of a unitary operation described by Hamiltonian dynamics was introduced in Ref.~\cite{pme},
where a randomization is applied to an auxiliary system and its initial state is prepared in the maximally mixed state.
Here we show that a similar effect can be implemented by applying randomization to the target system directly, 
instead of using an auxiliary system for the case of $d=2$.  
The main difference is that while the algorithm presented in Ref.~\cite{pme} performs a randomization on the auxiliary state,
our algorithm performs a randomization on the quantum operations.

While the introduction of a basis randomization comb is intended to apply to infinitesimal Hamiltonian dynamics, i.e., a unitary operation close to the identity operation for obtaining the approximate controllization of Hamiltonian dynamics, the definition of a basis randomization comb is valid for any quantum operation.
That is, the set of input operations to be neutralized is given by the set of an arbitrary quantum operation, 
but the error of the approximation depends on the input operations.
A generalization of the basis randomization comb for general $d$-dimensional systems is also straightforward.

\begin{figure}
\includegraphics[width=0.5\hsize]{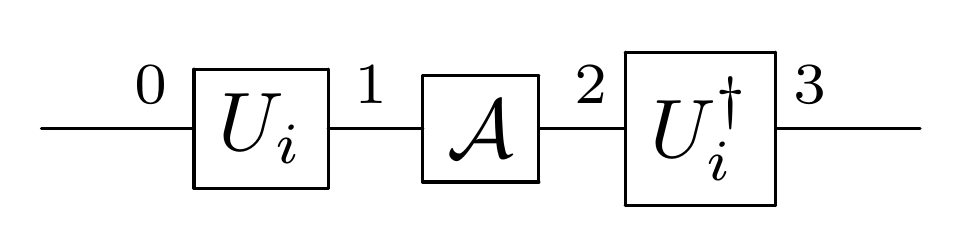}
\caption{Quantum circuit for the basis randomization comb.
The input quantum operation is ${\mathcal A}$, and the action of the basis randomization comb is given by applying a pair of unitary operations, ${U_i}$ randomly chosen from a set $\{ U_i \}$ and its inverse ${U_i^\dagger}$, before and after the quantum operation $\acal$.}\label{fig:randomization}
\end{figure}

Consider a quantum operation $\acal: L(\hcal_1 = \mathbb{C}^2) \to L(\hcal_2 = \mathbb{C}^2)$ whose Choi operator on $\hcal_1 \otimes \hcal_2$ is given by
\begin{align}
 J_\acal = \sum_{\alpha,\beta = I,X,Y,Z}  c_{\alpha,\beta} \kketbra{\alpha}{\beta}, \label{eq:def_choi_two_dim}
\end{align}
where $I, X, Y, Z$ represent the Pauli operators and $ c_{\alpha,\beta}$ is a coefficient.
We consider the basis randomization comb $\rcal_S$ implemented by the quantum circuit shown in Fig.~\ref{fig:randomization}  for a set of unitary operators $ R:= \{ U_i \}$.
The unitary operators composing the set $R$ are not necessary to be mutually orthogonal in general.  
We assume a unitary operator in the set is chosen uniformly randomly with probability $1/\abs{R}$ for simplicity.
We analyze the cases for two sets for $R$, a set consists of the Pauli operators in this appendix and another set consists of the Clifford operators in Appendix~\ref{ap:clifford_randomization}.
The action of $\rcal_R$ on $\acal$ denoted as $\ave{\acal}{R}$ is given by
\begin{align}
\ave{\acal}{R} = \frac{1}{\abs{R}} \sum_{U_i \in R} {U_i}^\dagger \circ \acal \circ {U_i},
\end{align}
which is implemented by applying a pair of unitary operations, ${U_i}$ randomly chosen from a set $\{ U_i \}$ and its inverse ${U_i^\dagger}$, before and after the quantum operation $\acal$.
The Choi operator of the basis randomization comb $\rcal_R$ is given by
\begin{align}
{\mathcal J}_{\rcal_R} = \frac{1}{\abs{R}}  \sum_{U_i \in R} \pproj{U_i}_{01} \otimes \pproj{U_i^\dagger}_{23}, \label{eq:def_randomize_comb_choi}
\end{align}
on $\hcal_0 \otimes \hcal_1 \otimes \hcal_2 \otimes \hcal_3$.
The Choi operator of the quantum operation transformed by the basis randomization comb ${\mathcal J}_{\rcal_R}$ for the input quantum operation $\acal$ is given by
\begin{align}
 J_{\ave{\acal}{R}} %
= \sum_{U_i \in R}\sum_{\alpha,\beta} c_{\alpha,\beta} \kketbra{U_i^\dagger\alpha U_i }{U_i^\dagger \beta U_i} \label{eq:def_random_channel}
\end{align}
on $\hcal_0 \otimes \hcal_3$. 
 
We investigate the action of the basis randomization comb with a set of the Pauli operators 
$R_P := \{ U_0=I,~U_1=X,~U_2=Y,~U_3=Z \}$ here.
A set of the Pauli operators $R_P$ forms a 1-design~\cite{design1,design2}. 
In Appendix~\ref{ap:clifford_randomization}, we analyze the basis randomization comb with a set of the Clifford operators $R_C$, which forms a 1-, 2- and 3- design~\cite{design1,design2} to investigate the difference caused by the sets of unitary operators used in the basis randomization comb.

For simplifying calculations, we introduce vectors ${\bi v}^{(i)}$ defined as
\begin{align}
 v^{(i)}_k : = 
 \begin{cases}
 1/2 ~~ (k=0,i) \\
 -1/2 ~~ ( \text{otherwise}).
 \end{cases} 
\end{align}
A set of vectors ${\bi v}^{(i)}$ forms an orthonormal basis of ${\mathbb R}^4$.  
The Pauli operators satisfy a commutation rule 
\begin{align}
 U_k U_i U_k = 4 \times v_k^{(i)} U_i.  \label{eq:pauli_commutation}
\end{align}
Since the Choi operator $ J_\acal$ can be decomposed in terms of the orthogonal basis $\{ \kket{U_i} \}$ as (equivalent to Eq.~\eqref{eq:def_choi_two_dim})
\begin{align}
J_\acal = \sum_{i,j} c_{i,j} \kketbra{U_i}{U_j},
\end{align}
the Choi operator of the transformed quantum operation $J_{\ave{\acal}{R_P}}$ is calculated to
\begin{align}
J_{\ave{\acal}{R_P}} &= \sum_{i,j}\sum_{k} v^{(i)}_k v^{(j)}_k c_{i,j} \kketbra{U_i}{U_j} \label{eq:calc_pauli_rand} \nonumber \\
 &= \sum_{i,j} {\bi v}^{(i)} \cdot {\bi v}^{(j)} c_{i,j} \kketbra{U_i}{U_j} \nonumber \\
 &= \sum_{i} c_{i,i} \kketbra{U_i}{U_i}.
\end{align}
Thus, the basis randomization comb with $R_P$ transforms the quantum operation $\acal$ to the quantum operation $\ave{\acal}{R_P}$ of which Choi operator is given by
\begin{align}
 J_{\ave{\acal}{R_P}}  = c_{00} \cdot J_{\rm id} + c_{11} \cdot J_\mathcal{X} + c_{2 2} \cdot J_\mathcal{Y} + c_{33} \cdot J_\mathcal{Z}, \label{eq:pauli_randomized}
\end{align}
where $\mathcal{X} ,\mathcal{Y}$ and $\mathcal{Z} $ denote the unitary operations by the Pauli operators $X,Y$ and $Z$, respectively, and $J_\mathcal{X}$, $J_\mathcal{X}$ and $J_\mathcal{Z}$ are the Choi operators of the corresponding Pauli operations.

We first consider a class of unitary operations given by infinitesimal Hamiltonian dynamics of a time-independent Hamiltonian $H$ as $\delta U = e^{-iH\delta t}$.
For a unitary operation $\delta \ucal$ described by a unitary operator $\delta U = e^{-iH\delta t} = I - i H \delta t + O(\delta t^2)$, the Choi operator of the transformed operation by the basis randomization comb $\jcal_{C_{\mathcal{R}_{R_P}}}$ is given by using Eq.~(\ref{eq:def_random_channel}) as
\begin{align}
J_{\ave{ \delta U}{R_P}} &= \pproj{I} + \sum_i (-i U_i^\dag H U_i \delta t) \pproj{I} \nonumber \\
&  + \sum_i \pproj{I} (-i U_i^\dag H U_i \delta t)^\dag + O(\delta t^2). \label{eq:motivate_pauli}
\end{align}
We see that the approximate neutralization for any unitary operation in this class with an error of $O(\delta t^2)$ is realized if the second and third terms in Eq.~\eqref{eq:motivate_pauli} vanish.

We further consider a quantum operation given by $U = e^{-iHt}$,
and apply the basis randomization comb with $R_P$ for each time interval $\delta t = t/n$ where $n$ is the number of division of the Hamiltonian dynamics in the duration time $t$.  

In this case, by considering $ \delta U = I - iH \delta t - H^2 \delta t^2 / 2 +  O( \delta t^3 )$ as the unitary operator for each time interval, 
we obtain $J_{\ave{\delta \ucal}{R_P}}$ with coefficients defined by Eq.~\eqref{eq:pauli_randomized} as
\begin{align}
c_{00} &= 1 + [ (\Tr H)^2 - d (\Tr H^2 ) ]  \delta t^2 / d^2 + O( \delta t^4 )  \nonumber \\
c_{11} &= ( \Tr HX )^2 \delta t^2 / d^2  + O( \delta t^4 ) \nonumber \\
c_{22} &= ( \Tr HY )^2 \delta t^2 / d^2  + O( \delta t^4 ) \nonumber \\
c_{33} &= ( \Tr HZ )^2 \delta t^2 / d^2  + O( \delta t^4 ) .
\end{align}
When the basis randomization comb is applied $n$ times, 
the resulting quantum operation is given by $( \langle {\delta U} \rangle_{S_P} )^n$.
Since any multiplication of Pauli operations results also a Pauli operation, 
the Choi operator of this operation can be decomposed in the form of Eq.~\eqref{eq:pauli_randomized}, namely,
\begin{align}
c^{(P)}_{0}  J_\mathrm{id} + c^{(P)}_{1} J_\mathcal{X} + c^{(P)}_{2} J_\mathcal{Y} + c^{(P)}_{3} J_\mathcal{Z}
\end{align}
with the coefficients
\begin{align}
c^{(P)}_{0} &= 1 + \frac{1}{n} [ (\Tr H)^2 - d (\Tr H^2 ) ] \frac{t^2}{d^2} \notag \\
 &\quad+ \frac{1}{2n^2} \{ [ (\Tr H)^2 - d (\Tr H^2 ) ]^2 \notag \\
&\quad + ( \Tr HX )^4 +  ( \Tr HY )^4 + ( \Tr HY )^4 ] \} \frac{t^4}{d^4} \notag \\
&\quad + O( \frac{1}{n^3} ) \label{eq:cc_pauli_coef_id} \\
c^{(P)}_{1} &= \frac{1}{n} ( \Tr HX )^2 \frac{t^2}{d^2} + O(\frac{1}{n^2}) \\
c^{(P)}_{2} &= \frac{1}{n} ( \Tr HY )^2 \frac{t^2}{d^2} + O(\frac{1}{n^2}) \\
c^{(P)}_{3} &= \frac{1}{n} ( \Tr HZ )^2 \frac{t^2}{d^2} + O(\frac{1}{n^2}).
\end{align}
Thus, for large enough $n$, the basis randomization comb with $R_P$ transforms any unitary operation generated by Hamiltonian dynamics to 
\begin{align}
\mathrm{id} + O(1/n),
\end{align}
which is close to the identity operation,
and thus this basis randomization comb is an approximate neutralization comb if it is applied to a unitary operation generated by Hamiltonian dynamics with a small enough interval $t/n$ $n$ times.

The controlled version of (a single element of) this basis randomization comb $\jcal_{C_{\mathcal{R}_{R_P}}}$ is determined by an operator $S_0$ as 
\begin{align}
\jcal_{C_{\mathcal{R}_{R_P}}} &=  \proj{00} \otimes \jcal_{\rcal_{R_P}} + \proj{11} \otimes \pproj{I} \notag \\
 &+ \ketbra{00}{11} \otimes \kketbra{S_0}{I} + \ketbra{11}{00} \otimes \kketbra{I}{S_0}.
 \end{align}
Note that the corresponding Kraus representation is given by $\{ \frac{1}{2} U_i \otimes U_i^\dag | U_i \in R_P\}$,
and $S_0$ is in the form of $S_0 = \frac{1}{2} \sum_i \alpha_i^\ast U_i \otimes U_i^\dag$ with $\sum_i \abs{\alpha_i}^2 = 1$.
The action of $\jcal_{\mathcal{R}_{R_P}}$ on  the Choi operator of an arbitrary unitary operation $\pproj{U}$ is 
\begin{align}
&\Tr_{12} [ \jcal_{\mathcal{R}_{R_P}} ( \pproj{U} )^T ]  \notag \\
&= \ketbra{00}{00} \otimes J_{\langle U \rangle_{R_P}}
+ \ketbra{11}{11} \otimes J_U  \notag \\
&\quad + \ketbra{11}{00} \otimes \bbra{U^*} \kketbra{I}{S_0} \kket{U^*} + h.c. ,
\end{align}
and  $\bbra{U^*} \kket{S_0}$ in the off-diagonal coherence term is evaluated as
\begin{align}
\bbra{U^*} \kket{S_0} &= \bbra{I}_{12} (I_1 \otimes U_2^T) \nonumber\\
&\quad \frac{1}{2}(\sum_i \alpha_i^\ast (U_i)_1 \otimes (U_i^*)_2) (\kket{I}_{01} \otimes \kket{I}_{23}) \nonumber\\
&= \frac{1}{2} \bbra{I} (\sum_i \alpha_i^\ast (U_i^\dag U U_i)_1 \otimes I_2) (\kket{I}_{01} \otimes \kket{I}_{23}) \nonumber\\
&= \frac{1}{2} \kket{\sum_i \alpha_i^\ast (U_i^\dag U U_i)}_{03}, \label{eq:pauli_rand_0011}
\end{align}
where the subscripts denote the indices  of the Hilbert spaces of the target system.

\begin{figure}
\includegraphics[width=0.5\hsize]{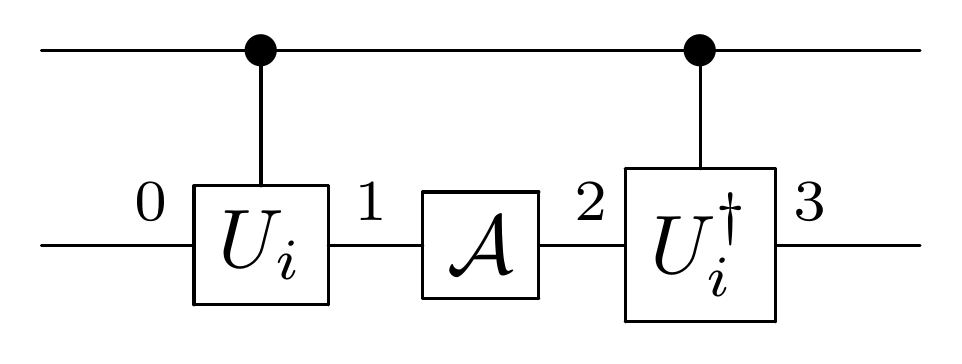}
\caption{A quantum circuit for the controlled  basis randomization comb.
The input quantum operation is ${\mathcal A}$, and the action of the  basis randomization comb is given by applying a pair of controlled unitary operations $\ccal_{U_i}$  chosen uniform randomly from a set $\{ U_i \}$ and its inverse $\ccal_{U_i}^\dag$ before and after the quantum operation $\acal$.
By repeating this circuit $n$ times,
the controlled unitary operation of a Hamiltonian dynamics $U=e^{-i H t}$ is implemented with an error of $O(1/n)$ with the global phase factor $\theta_U= (\mathrm{Tr} H/d)t$.
}
\label{fig:controlled_randomization}
\end{figure}

By requiring the  most coherently controlled identity operation  on the target system to be the identity operation  in the extended system including the control system, i.e., $I \mapsto \proj{0} \otimes I + \proj{1} \otimes I$, 
Eq.~\eqref{eq:pauli_rand_0011} should satisfy
\begin{gather}
\frac{1}{2} \kket{\sum_i \alpha_i^\ast (U_i^\dag I U_i)}_{03} = \kket{I}_{03}, \nonumber \\
\sum_i \alpha_i = 2.
\end{gather}
Thus, the coefficients are $\alpha_i = 1/2$ for all $i$,
and the operator $S_0$ is uniquely determined as
\begin{align}
S_0 = \frac{1}{4} (I \otimes I + X \otimes X + Y \otimes Y + Z \otimes Z),
\end{align}
and we obtain
\begin{align}
\bbra{U^*} \kket{S} &= \frac{1}{4} \kket{\sum_i (U_i^\dag U U_i)}_{03}.
\end{align}
A Kraus representation of this  most coherently controlled neutralization comb is given by 
\begin{align}
\{ &\proj{0} \otimes I \otimes I + \proj{1} \otimes I \otimes I, \nonumber\\
&\proj{0} \otimes X \otimes X + \proj{1} \otimes I \otimes I, \nonumber\\
&\proj{0} \otimes Y \otimes Y + \proj{1} \otimes I \otimes I, \nonumber\\
&\proj{0} \otimes Z \otimes Z + \proj{1} \otimes I \otimes I \},
\end{align}
and  one possible implementation in the quantum circuit is shown in Fig.~\ref{fig:controlled_randomization}.

When we apply the controlled  basis randomization comb $n$ times, since this comb does not change the state of the control qubit,
the term corresponding to Eq.~\eqref{eq:pauli_rand_0011} is evaluated as
\begin{align}
 I \otimes \frac{1}{4^n}(\sum_i U_i^\dag U U_i)^n \kket{I}.
\end{align}
For the case $U = e^{-iH\delta t} =  I - iH \delta t - H^2 \delta t^2 / 2 +  O( \delta t^3 )$ with $\delta t = t/n$,  we have
\begin{align}
&\frac{1}{4}(\sum_i U_i^\dag U U_i) \nonumber \\
&= I - i \delta t (\frac{1}{4} \sum_i U_i^\dag H U_i)  \nonumber \\
&\quad - \frac{1}{2} \delta t^2 ( \frac{1}{4} \sum_i U_i^\dag H^2 U_i) + O( \delta t^3 ) \nonumber \\
&= I - i\delta t (\Tr H) I / d - \frac{1}{2} \delta t^2 ( \Tr H^2) I/d + O(\delta t^3),
\end{align}
and we obtain
\begin{align}
\frac{1}{4^n}(\sum_i U_i^\dag U U_i)^n = e^{-i (\Tr H / d) t} I + O(1/n).
\end{align}

Finally,  we obtain the Choi operator of the quantum operation transformed from  $U = e^{-iHt}$ by the controlled  basis randomization comb with $R_P$ as
\begin{align}
&\ketbra{00}{00} \otimes J_I
+ \ketbra{11}{11} \otimes J_U \\
&\quad +  \ketbra{11}{00} \otimes \kketbra{e^{i (\Tr H / d) t} U}{I} + h.c.
+ O(1/n),
\end{align}
which converges to the  (fully coherently) controlled unitary operation, that is $C_{e^{-iHt}} = \proj{0} \otimes I + \proj{1}  \otimes e^{i (\Tr H /d) t} e^{-iHt}$ in the limit of $n \to \infty$.

\section{Basis randomization with Clifford operators}\label{ap:clifford_randomization}

In Appendix~\ref{ap:pauli_randomization}, we considered the basis randomization using  a set of the Pauli operators $R_P$,
 for aiming to ``nullify'' the terms of $O(\delta t)$ in Eq.~\eqref{eq:motivate_pauli}.
 The Choi operator of the transformed operation by the basis randomization comb given by Eq.~\eqref{eq:def_random_channel} is of the form $\sum_i (U_i^\dag \otimes U_i^T) \jcal_\acal (U_i^\dag \otimes U_i^T)^\dag$.
Since the Choi operator of the identity operation, $\pproj{I}$, is the fixed point of $\int dU (U^\dag \otimes U^T) \cdot (U^\dag \otimes U^T)^\dag$,
it is expected that this integral transforms any Choi operator approximately to $\pproj{I}$.

The corresponding effect can be achieved for $\sum_i (U_i^\dag \otimes U_i^T) \cdot (U_i^\dag \otimes U_i^T)^\dag$ if we choose the set $S = \{ U_i \}$ to be a 2-design (by definition of 2-design)~\cite{design2}.
Thus, the basis randomization by a 2-design may behave better than a 1-design, that is, the Pauli randomization.
In this appendix, we  analyze the basis randomization comb employing a 2-design using a set of the Clifford operators of a 1-qubit system.

We first summarize the properties of the Clifford group that we use in the following~\cite{clifford1}.
Clifford group  $G_C$ is the group of operators whose conjugation transforms any Pauli operator in the Pauli group $G_P$  into another Pauli operator, that is, 
\begin{align}
 \forall\, U \in G_P, ~ \forall\, V \in G_C, ~ V U V^\dagger \in G_P.
\end{align} 
Note that the Clifford group has a trivial center $Z$
\begin{align}
Z = \{ \pm I, \pm i I, \pm e^{i\frac{\pi}{4}} I, \pm e^{3i\frac{\pi}{4}} I \},
\end{align}
whose elements can only change the global phase.
In the density operator formalism, any unitary operator  representing a unitary operation appears together with its complex conjugate, and  thus the effect of the global phase is always canceled.
Thus, we only consider $R_C$, the residue class of $G_C$ divided by $Z$, that is, $R_C := G_C / Z$.  
Since the set of the Pauli operators $R_P$ is a normal subgroup of $R_C$,
we can define the residue group $R_{C/P} := R_C/G_P$.

The representative elements of $R_{C/P}$ are given by the following six operators $R_{C/P} = \{ V_\sigma \}$, where $\sigma$ is a permutation among $\{1,2,3\}$,
\begin{align}
V_{\mathrm{id}} &= I \\ 
 V_{(1,2)} &= \left( \begin{matrix} 1 & 0 \\ 0 & i \end{matrix} \right) \\ 
 V_{(2,3)} &= \frac{1}{\sqrt{2}}\left( \begin{matrix} e^{i \frac{\pi}{4}} & e^{-i\frac{\pi}{4}} \\ e^{-i\frac{\pi}{4}} & e^{i\frac{\pi}{4}} \end{matrix} \right) \\
 V_{(3,1)} &= \frac{1}{\sqrt{2}}\left( \begin{matrix} 1 & 1 \\ 1 & -1 \end{matrix} \right) \\
 V_{(1,2,3)} &= \frac{1}{\sqrt{2}}\left( \begin{matrix} 1 & i \\ 1 & -i \end{matrix} \right) \\
 V_{(3,2,1)} &= \frac{1}{\sqrt{2}}\left( \begin{matrix} 1 & 1 \\ i & -i \end{matrix} \right).
\end{align} 

The Choi operator $J_{\langle \acal \rangle_{R_C}}$ of the quantum operation $\acal$  transformed by the basis randomization comb with the set of the Clifford operators $R_C$ is given by
\begin{align}
 &J_{\langle \acal \rangle_{R_C}} \notag \\
& =  \frac{1}{24}  \sum_{\alpha,\beta}\sum_{U_i \in R_P} \sum_{V_j \in R_{C/P}} c_{\alpha,\beta} \kket{U_i^\dagger V_j^\dagger \alpha V_j U_i}\bbra{U_i^\dagger V_j^\dagger \beta V_j U_i}.
\end{align}
 Since $V_j^\dagger \alpha V_j$ and $V_j^\dagger \beta V_j$ are Pauli operators by definition of the Clifford operators, we obtain 
\begin{align}
 J_{\langle \acal \rangle_{R_C}} =  \frac{1}{6}  \sum_{\alpha} \sum_{V_j \in R_{C/P}} c_{\alpha,\alpha} \kket{V_j^\dagger \alpha V_j }\bbra{V_j^\dagger \alpha V_j}
\end{align}
 similarly to the calculation in Eq.~\eqref{eq:calc_pauli_rand}.
Note that the following formula holds for $\alpha = U_i$  for $i=0,1,2,3$,
\begin{align}
\kket{V_\sigma U_i V_\sigma^\dagger }\bbra{V_\sigma U_i V_\sigma^\dagger} = \kket{U_{\sigma(i)} }\bbra{U_{\sigma(i)}},
\end{align}
where we  set $\sigma(0):=0$.
This means that the matrix element spanned by $\kket{X},\kket{Y},\kket{Z}$ is completely mixed by Clifford operations.  
Therefore, we obtain
\begin{align}
 J_{\langle \acal \rangle_{R_C}} = c_{00} \cdot J_{\rm id} + \frac{c_{11}  + c_{22} + c_{33}}{3}  \left( J_\mathcal{X} +J_\mathcal{Y} + J_\mathcal{Z} \right).
\end{align}
By using the depolarizing channel ${\mathcal D}:= ({\rm id} + \mathcal{X} +\mathcal{Y} + \mathcal{Z} )/4$,  the Choi operator of the transformed operation is also  represented as
\begin{align}
  J_{\langle \acal \rangle_{R_C}} &= \left(c_{00} -\frac{c_{11}  + c_{22} + c_{33}}{3} \right) J_\mathrm{id} \nonumber \\
 &\quad + \frac{4}{3} \left(c_{11}  + c_{22} + c_{33}  \right) J_{\mathcal D}.
\end{align}

Similarly to the case of the basis randomization comb with $R_P$,
we consider the quantum operation given by a time-independent Hamiltonian $H$, i.e., $U = e^{-iHt}$,
and apply the  basis randomization comb for each time interval $\delta t = t/n$.
When the  basis randomization  with $R_C$ is applied $n$ times, 
 the Choi operator of the transformed operation is given by 
\begin{align}
c^{(C)}_{0}  J_\mathrm{id} + c^{(C)}_{1} J_\mathcal{X} + c^{(C)}_{2} J_\mathcal{Y} + c^{(C)}_{3} J_\mathcal{Z}
\end{align}
with the coefficients
\begin{align}
c^{(C)}_{0} &= 1 + \frac{1}{n} [ (\Tr H)^2 - d (\Tr H^2 ) ] \frac{t^2}{d^2} \notag \\
&\quad + \frac{1}{2n^2} \{[ (\Tr H)^2 - d (\Tr H^2 ) ]^2 \notag \\
&\quad+ \frac{1}{3}[( \Tr HX )^2 + ( \Tr HY )^2 + ( \Tr HY )^2]^2 \} \frac{t^4}{d^4} \notag \\
&\quad + O( \frac{1}{n^3} ) \label{eq:cc_clifford_coef_id}
\end{align}
\begin{align}
&c^{(C)}_{1} = c^{(C)}_{2} = c^{(C)}_{3} \notag \\
&\quad= \frac{1}{3n} [ ( \Tr HX )^2 + ( \Tr HY )^2 + ( \Tr HY )^2 ] \frac{t^2}{d^2} + O(\frac{1}{n^2}).
\end{align}
The coefficient of  $J_\mathrm{id}$ coincides with that  for the case with $R_P$ up to the order $1/n$. 
However, the basis randomization comb with $R_C$ performs worse than the case with $R_P$ in the sense that the coefficient of $J_\mathrm{id}$ is smaller, when the terms of $O(1/n^2)$ are considered.
The basis randomization comb with $R_C$  transforms any unitary operation  generated by Hamiltonian dynamics to 
\begin{align}
\mathrm{id} + O(1/n),
\end{align}
which is close to the identity  operation with error of $O(1/n)$, and thus this  basis randomization comb is an approximate neutralization comb.

The  most coherently controlled version of (a single element of) the basis randomization comb $\jcal_{C_{\mathcal{R}_{R_C}}}$ is determined by an operator $S_0$ similarly to the case of $R_P$.
The Kraus representation of the basis randomization comb with $R_C$ is given by $\{ \frac{1}{\sqrt{24}} U_i \otimes U_i^\dag | U_i \in R_C\}$.  Since the dimension of the linear span of $\mathcal{L}(\mathbb{C}^2 \otimes \mathbb{C}^2)$ is 16 while this set contains 24( $> 16$) elements, this set of operators is over-complete. 
Thus we need to find an orthogonal Kraus representation.  The span of  the Kraus representation is invariant under the swap operation $U_{swap}$ between the first and the second Hilbert space,
because any element is of the form $K = \sum_i \alpha_i U_i \otimes U_i^\dag$ with $U_i \in R_C$,
and $U_{swap} K U_{swap}$ is also in the span.
Thus, the span is in the $d(d+1)/2 = 10$ dimensional symmetric subspace.
By calculating the spectral decomposition of the Choi operator corresponding to this Kraus representation, 
we  can check that the Kraus operators actually span the 10-dimensional symmetric subspace.
Specifically, the Kraus representation is given by
\begin{align}
\{ S_0 = \frac{1}{4} (I \otimes I + X \otimes X + Y \otimes Y + Z \otimes Z), S_1, \ldots, S_9 \},
\end{align}
where $ \{ S_i \}$  is a set of orthogonal operators in the symmetric subspace 
 satisfying $\Tr S_i^\dag S_i = 1/3$ for $i=1,\ldots,9$.
Note that $\Tr S_0^\dag S_0 = 1$.
Thus, the  off-diagonal coherent term of the  most coherent controlled comb is characterized by 
\begin{align}
S_0 = \frac{1}{4} (I \otimes I + X \otimes X + Y \otimes Y + Z \otimes Z),
\end{align}
which coincides with the case with the basis randomization with Pauli operators\footnote{
Precisely, the global phase is not uniquely determined by maximizing the Hilbert-Schmidt norm of the operator, 
and $e^{i\theta} S_0$ for any real parameter $\theta$ is also a candidate instead of $S_0$.
We choose $\theta = 0$ by requiring the  most coherently controlled version of the identity operation to be also the identity operation, i.e., $I \mapsto \proj{0} \otimes I + \proj{1} \otimes I$.
}.
This indicates that the maximum off-diagonal coherent terms are the same for controllization of the basis randomization comb with $R_P$ and $R_C$ for up to $O(1/n)$ approximation.

Since the  basis randomization comb  with $R_C$ does not behave better than the one with $R_P$ for Hamiltonian dynamics with the terms of up to $O(1/n)$, and the coherent terms of both cases coincide, we conclude that using the Clifford operators does not improve controllization of Hamiltonian dynamics. 
Moreover,  the analysis of the terms with $O(1/n^2)$ shows that the performance of the basis randomization with $R_C$ as approximate neutralization turns out to be worse than that of $R_P$ in general. 
Thus, it is enough to use $R_P$ for the task of controllization of Hamiltonian dynamics using the most coherently controlled basis randomization comb.

\section{Kraus representation for quantum combs}\label{ap:comb_kraus}

In this appendix, we derive the conditions for an $N$-slot quantum comb ${\mathcal S} : L( \hcal_1 \otimes \hcal_2 \otimes \cdots \otimes \hcal_{2N} ) \to L( \hcal_0 \otimes \hcal_{2N+1})$ in terms of its Kraus representation, instead of the Choi representation.
Let $\{ S_i \}$ be the Kraus representation of ${\mathcal S}$, that is,
\begin{align}
 {\mathcal S} \left[ J \right] = \sum_i S_i J S^\dagger_i,
\end{align}
with $S_i : \hcal_1 \otimes \dots \otimes \hcal_{2N} \rightarrow \hcal_0 \otimes \hcal_{2N+1}$.
Since the complete positivity condition of a quantum comb given by Eq.~\eqref{eq:condition_comb_cp} is automatically satisfied by the form of the Kraus representation, 
the only remaining condition to be derived is the condition given by Eq.~\eqref{eq:condition_seq}.

We first consider the condition given by Eq.~\eqref{eq:condition_seq}  for $k=N$, that is,
\begin{align}
\Tr_{2N+1} {\mathcal J}_{\mathcal S} = \Tr_{2N,2N+1} {\mathcal J}_{\mathcal S} \otimes \frac{I_{2N}}{d_{2N}}, \label{eq:condition_seq_last}
\end{align}
where ${\mathcal J}_{\mathcal S}$ is the Choi operator of ${\mathcal S}$.
This condition is equivalent to
\begin{align}
&\Tr_{0;2N-1} (A_0 \otimes B_{1;2N-1} \otimes I_{2N}) \Tr_{2N+1} {\mathcal J}_{\mathcal S} \notag \\
&= \Tr_{0;2N-1} (A_0 \otimes B_{1;2N-1} \otimes I_{2N}) \Tr_{2N,2N+1} {\mathcal J}_{\mathcal S} \otimes \frac{I_{2N}}{d_{2N}} \notag \\
&= c \cdot I_{2N}
\label{normalizationcon}
\end{align}
where $\Tr_{0;2N-1}$ denotes the partial trace taken over $\hcal_0 \otimes\hcal_1 \otimes \cdots \otimes \hcal_{2N-1}$,  holding for any $A_0 \in L(\hcal_0)$ and $B_{1;2N-1} \in L(\hcal_1 \otimes \cdots \otimes \hcal_{2N-1})$ and a complex number depending on $A_0,B_{1;2N_1}$.

By rewriting the Choi operator ${\mathcal J}_{\mathcal S}$ in terms of the Kraus operators $\{ S_i \}$, the equality
\begin{align}
&\Tr_{0;2N-1} (A_0 \otimes B_{1;2N-1} \otimes I_{2N}) \Tr_{2N+1} {\mathcal J}_{\mathcal S} \notag\\
&= \sum_{k,k'} \ketbra{k}{k'}_{2N} \cdot \Tr_{2N} \big[ \ketbra{k}{k'}_{2N} \notag \\
&\quad \times \Tr_{1;2N-1} (B_{1;2N-1} \otimes I_{2N} ) \sum_i  S_i^\dagger (A_0 \otimes I_{2N+1} ) S_i \big],
\end{align}
 holds for any $A_0 \in L(\hcal_0)$ and $B_{1;2N-1} \in L(\hcal_1 \otimes \cdots \otimes \hcal_{2N-1})$. 
Thus, we obtain the following condition in terms of the Kraus operators $\{ S_i \}$:
For any linear operators $A_0 \in L(\hcal_0)$ and $B_{1;2N-1} \in L(\hcal_1 \otimes \cdots \otimes \hcal_{2N-1})$, the following equality holds
\begin{align}
 \partialtr{1;2N-1}{ (B_{1;2N-1} \otimes I_{2N} ) \sum_i  S_i^\dagger (A_0 \otimes I_{2N+1} ) S_i} \notag\\
 = c \cdot I_{2N}, \label{eq:condition_seq_kraus_last}
\end{align}
where $c$ is a complex number given by the trace of the l.h.s. divided by $d_{2N}$.

\begin{figure}
\includegraphics[width=0.85\linewidth]{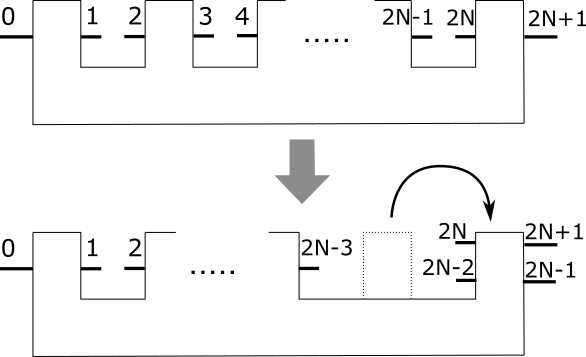}
\caption{(Above) An $N$-slot quantum comb $\mathcal{S}$, an abstract description of a quantum circuit which calls $N$ quantum operations  ${\mathcal A}_k$ for $k=1,2, \cdots , N$.
(Below) The ($N-1$)-slot quantum comb $\mathcal{S}^{(1)}: \hcal_1 \otimes \hcal_2 \otimes \cdots \otimes (\hcal_{2N-2} \otimes \hcal_{2N}) \rightarrow \hcal_0 \otimes \left( \hcal_{2N-1} \otimes \hcal_{2N+1} \right)$ 
induced from the $N$-slot quantum comb $\mathcal{S}$.
 The ($N-k$)-slot quantum comb ${\mathcal S}^{(k)}$ for $k=1,2,\cdots, N-1$ is defined by repeating this procedure.}\label{fig:comb_reduction}
\end{figure}

 An $N$-slot quantum comb $\mathcal{S}$ can be redefined as an ($N$-1)-slot quantum comb denoted by  ${\mathcal S}^{(1)}: L(\hcal_1 \otimes \cdots \otimes \hcal_{2N-4} \otimes \hcal_{2N-2}^\prime) \rightarrow L(\hcal_0 \otimes \hcal_{2N-1}^\prime)$ with $\hcal_{2N-2}^\prime = \hcal_{2N-2} \otimes \hcal_{2N}$ and $\hcal_{2N-1}^\prime = \hcal_{2N-1} \otimes \hcal_{2N+1}$, as shown in Fig.~\ref{fig:comb_reduction}.  
The corresponding Kraus representation $\{ S_i^{(1)} \}$ is given as
\begin{align}
 S_i^{(1)}  = S_i \kket{I_{2N-1}}_{2N-1,2N-1},
 \end{align}
 where $S_i^{(1)}$ can be understood as the operator $S_i$ with its domain $\hcal_{2N-1}$ been moved to the range.
Recursively, we can define an ($N$-$k$)-slot quantum combs ${\mathcal S}^{(k)}$,
and its Kraus representation $\{ S_i^{(k)} \}$ of ${\mathcal S}^{(k)}$ is given as
\begin{align}
 S_i^{(k)}  = S_i \kket{I_{2N-2k+1}} \otimes \kket{I_{2N-2k+3}} \otimes \cdots \otimes \kket{I_{2N-1}}.
\end{align} 
We also set $\mathcal{S}^0:=\mathcal{S}$ and $S_i^{(0)}:=S_i$.

Then the condition Eq.~\eqref{eq:condition_seq_kraus_last} can be transformed to the condition for ${\mathcal S}^{(k)}$ with $k = 0, 1,\ldots, N-1$, which correspond to the condition Eq.~\eqref{eq:condition_seq} with $N-k$, as shown in the following.
Note that Eq.~\eqref{eq:condition_seq_kraus_last} corresponds to the case of $k=0$.
For all liner operators $A_0$ and $B_{1;2N-2k-1}$, there is a complex number $c$ such that 
\begin{multline}
 \Tr_{1;2N-2k-1}
 \Bigg( (B_{1;2N-2k-1} \otimes I_{{\mathcal H}^{(k)}} ) \notag \\
 \times  \sum_i  S_i^{(k)\dagger} (A_0 \otimes I_{{\mathcal K}^{(k)}}) S_i^{(k)} \Bigg) %
  = c \cdot I_{{\hcal}^{(k)}} , \label{eq:condition_seq_kraus}
\end{multline}
where 
\begin{align}
 {\hcal}^{(k)} &= \bigotimes_{l=0}^{k} \hcal_{2N-2l}, \\
 {\mathcal K}^{(k)} &= \bigotimes_{l=0}^{k} \hcal_{2N-2l+1}.
\end{align}
The remaining condition given by Eq.~\eqref{eq:condition_seq} is the one for $k=0$, 
and it is equivalent to
\begin{align}
\Tr_{\kcal^{(N)}} {\mathcal J}_{\mathcal S} = I_{\hcal^{(N)}}, 
\end{align}
when all other conditions given by Eq.~\eqref{eq:condition_seq} are satisfied.
Since this condition is similar to the trace-preserving condition of a map from $\hcal^{(N)}$ to $\kcal^{(N)}$,
it can be written with $\bbra{I_0} S_i^{(N)}$ as $\sum_i S_i^{(N)\dag} \pproj{I_0} S_i^{(N)} = I_{\hcal^{(N)}}$,
equivalently, we obtain
\begin{align}
\Tr_{\kcal^{(N)}} \sum_i  S_i^\dag (\pproj{I_0}_{0,0} \otimes I_{2N+1}) S_i = I_{\hcal^{(N)}}. \label{eq:condition_seq_kraus2}
\end{align}

As the complete positivity of the quantum comb Eq.~\eqref{eq:condition_comb_cp} is automatically satisfied, 
the condition for quantum comb in terms of the Kraus representation is given by Eq.~\eqref{eq:condition_seq_kraus} with $k=0,1,\ldots, N-1$ and Eq.~\eqref{eq:condition_seq_kraus2}.

\end{document}